\documentclass[9pt,shortpaper,twoside,web]{ieeecolor}
\usepackage{apg}
\makeatletter\let\NAT@parse\undefined\makeatother 
\usepackage[numbers]{natbib}
\usepackage{amsmath,amssymb,amsfonts}
\usepackage{algorithmic}
\usepackage{graphicx}
\usepackage{textcomp}
\usepackage{hyperref}

\def\BibTeX{{\rm B\kern-.05em{\sc i\kern-.025em b}\kern-.08em
    T\kern-.1667em\lower.7ex\hbox{E}\kern-.125emX}}
\usepackage[switch]{lineno}
\usepackage{subcaption}
\usepackage{booktabs,multirow,accents}

\usepackage{bm}\renewcommand{\vec}{\bm}
  
\makeatletter\def\operator@font{\sf}\makeatother

\DeclareMathOperator*{\argmax}{arg\,max}

\usepackage{lastpage,fancyhdr}
\usepackage{ifthen,hyperref}
\usepackage{lipsum}
\fancyheadoffset[L]{0.25in}
\fancyfootoffset[L]{0.25in}
\title{Visually Guided Swarm Motion Coordination via Insect-inspired Small Target Motion Reactions}
\author{Md Arif Billah, and Imraan A. Faruque
\thanks{M.A.~Billah is with Oklahoma State University, Stillwater, OK 74078 USA (e-mail: arif.billah@okstate.edu). }
\thanks{I.~A.~Faruque is with Oklahoma State University, Stillwater, OK 74078 USA (e-mail: i.faruque@okstate.edu).}
\thanks{This work was supported in part by ONR Young Investigator Award N00014-19-1-2216. Copyright by the authors June 2022; this work may be under consideration for publication and copyright may be transferred without notice, after which this version may no longer be accessible.}
}

\begin{document}
\maketitle

\begin{abstract}
Despite progress developing experimentally-consistent models of insect in-flight sensing and feedback for individual agents, a lack of systematic understanding of the multi-agent and group performance of the resulting bio-inspired sensing and feedback approaches remains a barrier to robotic swarm implementations. This study introduces the small-target motion reactive (STMR) swarming approach by designing a concise engineering model of the small target motion detector (STMD) neurons found in insect lobula complexes. The STMD neuron model identifies the bearing angle at which peak optic flow magnitude occurs, and this angle is used to design an output feedback switched control system. A theoretical stability analysis provides bi-agent stability and state boundedness in group contexts. The approach is simulated and implemented on ground vehicles for validation and behavioral studies. The results indicate despite having the lowest connectivity of contemporary approaches (each agent instantaneously regards only a single neighbor), collective group motion can be achieved. STMR group level metric analysis also highlights continuously varying polarization and decreasing heading variance.
\end{abstract}
\section{Introduction} \label{sec:intro}
Insects robustly achieve the coordination required for collective group flight, largely by leveraging a small collection of neurons in their visuomotor system. Tethered insect studies have identified individual neurons in the visual sensory pathway and their respective roles in environmental navigation. Despite the progress in identifying  the insect visuomotor neural pathways, the resulting flight control functions of these neural processes are not always known, particularly when stimulated by large numbers of neighboring agents.

Some insects demonstrate high visual acuity during interactions with prey, predators or conspecific neighbors. This interaction requires visual detection of targets in unstructured environments \citep{lingenfelter2021insect}.  Targets often subtend only a small region of the visual field relative to sensor resolution. The small-target motion detector (STMD) neurons found in the lobula complex of certain insects such as dragonflies, blowflies and hoverflies are highly sensitive to small, dark, fast-moving objects. Models predicting the behavior of STMDs suggest that these neural models can be used for target detection and pursuit in artificial vision-based robotic applications \citep{bagheri2017autonomous,bagheri2017performance,wiedermann2013biologically,wang2021time}. Although recent progress in multi-agent modeling has shown how a parallel wide-field visual processing pathway can participate in swarm motion \citep{humbert2005sensorimotor,billah2020bioinspired}, there is not a clear understanding for how the STMDs behave in a multi-agent environment. This study investigates whether the responses of STMD neurons can help insects participate in swarm motion through theory, simulation, and experiment.

This paper builds on \citet{billah2022multi}, which introduced the mathematical feedback model form and provided the first simulations of various individual feedback laws, and on \citet{billah2022modeling}, which suggests a theoretical bound on trajectories under average dwell time constraints for various individual feedback laws, without any further theoretical or trajectory analysis or experimental implementation. This paper builds on the general feedback approach identified in these papers by providing explicit theoretical convergence analysis of the behavior, which is applied to design a controller convergent to local formations. This study also formalizes the STMR approach's implicit visual interaction between agents through a graph theoretic representation, which supports understanding of the implicit information flow and instantaneous computational/attentional workload of this approach relative to alternatives. Finally, this study develops an experimental robotic implementation of the STMR approach, using ground robots to provide experimental data that quantifies the real-world performance variation of these visually-guided algorithms. This study also addresses a broader context by comparing the performance of this controller against existing swarm approaches. The addition of a stability proof, a graph-theoretic implicit information framework, development of an implemented robotic experiment and its data analysis, and contemporary swarm approach simulations and systematic comparison analysis, this study provides a more comprehensive treatment that motivated this paper.

The main contributions of this paper include (a) a concise retina-to-motor neuron dynamic model of STMD neuronal feedback, which goes beyond the STMD neuronal models as simply sensory mechanisms, (b) analysis and stability proof of state boundedness over time, necessary for reliable robotic implementations, (c) simulation and analysis of information flow showing low levels of connectivity, indicating low attentional requirements relative to contemporary approaches, and (d) experimental work using external recording of onboard implementation in robotic hardware, used to provide an uncertainty bound by showing repetitive behavior.

This paper is structured as follows: section~\ref{sec:lit_rev} reviews previous work related to the characteristics of small-target motion detector neurons, feedback laws for pursuit flight in insects and stability of switched systems. Section~\ref{sec:methods} proposes a mathematical model of the STMD neurons, designs a continuous-time multi-agent switched feedback control system using the STMD neuron model as a sensory function and analyzes stability of the designed multi-agent system. This section also discusses the simulation and experimental methods used for validation and further analysis. Section~\ref{sec:results&discussion} presents results and discusses the findings and outcomes from this study and their implications.  

\section{Previous Work and Background} \label{sec:lit_rev}
\subsection{Small target motion detector neurons}
The small-target motion detector (STMD) neurons were first identified in dragonflies \citep{o1993feature} by O'Carroll. STMDs have also been studied in dipteran flies such as hoverflies \citep{nordstrom2006insect} and blowflies \citep{o2014contrast}. Studies by \citet{nordstrom2006insect} have shown the STMDs to have extremely high contrast sensitivity and background motion rejection \citep{nordstrom2006insect}. Experimental studies found STMDs to be highly-sensitive to small moving targets \citep{wiederman2011discrimination}, exhibit selective attention to a particular target in the presence of multiple targets \citep{wiederman2013selective}, and show predictive gain modulation to target trajectories \citep{wiederman2017predictive}. The STMD neurons also perform gain modulation for low contrast targets \citep{geurten2007neural}, even amongst cluttered dynamic backgrounds \citep{evans2022dragonfly}, and the gain modulation response can depend on factors such as target angular size and velocity \citep{fabian2019properties}. \citet{lancer2022preattentive} further studied the selective attention property of STMDs that randomly tracks a particular target in the presence of multiple targets and discovered the ``lock on" mechanism where the STMD neurons fixate on a low-contrast target to reduce switching between targets even when a high-contrast distractor is introduced \citep{lancer2019target}, suggesting a varied mechanism for processing the visual stimuli compared to spatial wide-field integration (WFI) of optic flow \citep{krapp2022optic} performed in the lobula complex of the insect visual ganglia.  Although individual insect targets may be isolated and tracked by STMD neurons, the question of whether STMDs function to inhibit or enable conspecific swarm flight is not yet fully addressed in literature.  

Several quantitative models of STMDs have been proposed in recent years to define their underlying mechanisms \citep{wiedermann2013biologically, wang2021time, billah2022modeling, wiederman2008model, wang2018directionally, wang2019robust}. These include elementary STMD (ESTMD) \citep{wiederman2008model}, cascaded EMD-ESTMD \citep{wiedermann2013biologically}, ESTMD with feedback \citep{wang2021time}, directionally selective STMD (DSTMD) \citep{wang2018directionally} and STMD+ \citep{wang2019robust}. Implementation of some of these models on autonomous robots for target tracking has highlighted the effectiveness of these models in artificial visual feedback systems \citep{bagheri2017performance,bagheri2017autonomous}. 

\subsection{Insect individual feedback control}
Several individual feedback guidance laws have been used to describe the target tracking behavior of insects \citep{varennes2020two, strydom2017uas, mischiati2015internal, srinivasan2011visual, justh2006steering}. An engineering history underlies the comparatively straightforward pure pursuit algorithm \citep{shneydor1998missile}, in which the error angle defined as the difference between the pursuer heading and the bearing angle of the evader is reduced to zero. If the pursuer regulates its heading to a certain non-zero value of error angle instead, that feedback law is referred to as biased pursuit or deviated pure pursuit \citep{shneydor1998missile}. A proportional navigation controller aims to maintain a constant bearing angle of the evader \citep{peppas1992d}, and a hybrid controller is a combination of both the pure pursuit controller and proportional navigation controller. Motion camouflage strategies minimize relative motion to reduce the visual signature of the agent, and one method to achieve visual signature reduction is to regulate the optic flow generated by the pursuer on the retina of the evader to zero \citep{justh2006steering}. This approach achieves target tracking while hiding pursuer motion from the evader. The CDMC algorithm maintains constant distance to the evader while concealing the pursuer's motion, for simple linear trajectories of the evader \citep{strydom2017uas}. Effect of sensory noise and delay in the motion camouflage guidance law have also been considered for a more biologically consistent approach \citep{raju2016motion}. Studies suggest each of these feedback control strategies have been found in various insects \citep{varennes2020two}. Implementation of insect guidance laws for collision avoidance and pursuit on robotic platforms have shown consistency with theoretic formulations for individual agents \citep{huang2019biohybrid,colonnier2019bio}. Bio-inspired insect feedback laws are chosen as feedback for the multi-agent switched system developed in this paper that utilizes the response of STMD neurons for feedback control.

\subsection{Switched Systems}
A switched system is defined as a family of dynamical systems wherein a switching signal activates only one subsystem at a time. Switched systems have been widely used in various fields such as power electronics \citep{vasca2012dynamics}, automotive control \citep{wu2018optimal}, robotics \citep{namaki2012target} and aerial systems \citep{carrillo2013quad}. Significant effort has been directed toward developing stability and robustness criteria for switched systems. Stability of switched linear systems was studied in terms of construction of special classes of Lyapunov functions \citep{narendra1994common,mancilla2020uniform} which decreased over time despite switching to ensure stability. \citet{hespanha1999stability} showed that the stability properties of individual subsystems can be translated to the switched system when switching is sufficiently slow, i.e., the switching signal satisfies an average dwell-time bound. The use of average dwell time for stability was further investigated in \citep{vu2007input} to study the input-to-state stability (ISS) of switched continuous systems; in \citep{chen2016relaxed} to relax the ISS requirement on each subsystem for switching under disturbances; in \citep{alpcan2010stability,dorothy2016switched} for switched systems with multiple equilibria; and, more recently, in \citep{veer2019switched} for switched systems with multiple equilibria under disturbances. The stability criteria based on average dwell time constraints from \citet{veer2019switched} is used for stability analysis of the STMD-based multi-agent switched control system developed in this study. 

\section{Methods and Approach} \label{sec:methods}
In this section, the modeling, analysis and validation methods for the multi-agent feedback control system is described. A simplified model of the STMD neuron function is proposed in \ref{ssec:stmd_model}, a multi-agent switched feedback control system using the STMD function is developed in \ref{ssec:feedback_law}, stability analysis of the designed control system is provided in \ref{sec:stability_proof} followed by implementation in simulation (\ref{ssec:simulation}) and on ground robots (\ref{ssec:experiment}). 

\subsection{Small-Target Motion Detector (STMD) Neuronal Model} \label{ssec:stmd_model}
\begin{figure}\centering\includegraphics[width=0.8\columnwidth]{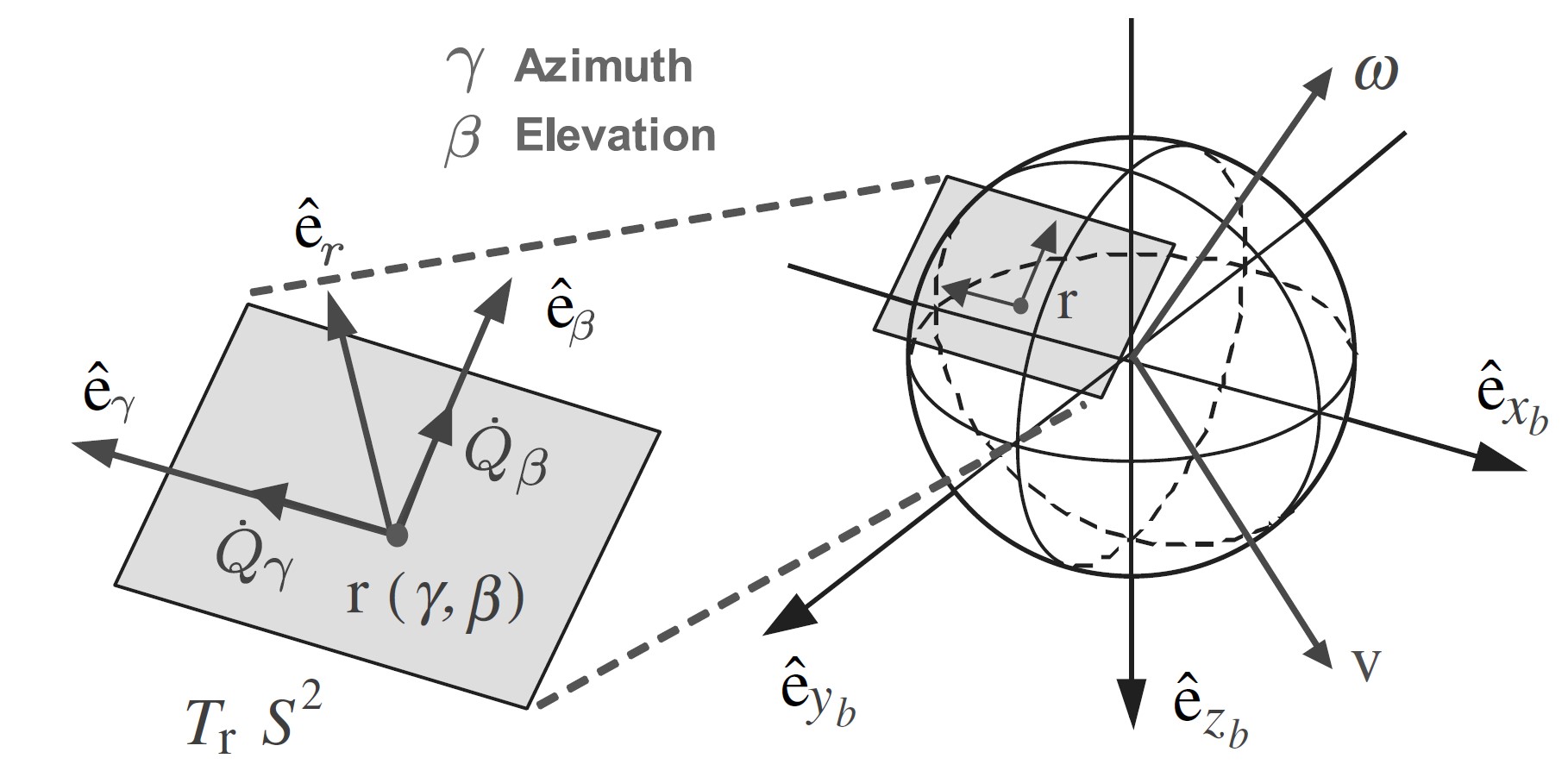}\caption{Geometry of the optic flow definition \citep{Humbert2005}.}\label{f:azimuthElevation}\end{figure}

The magnitude of optic flow signal depends upon the size, relative velocity and the distance between the observer and the target, along with the observer's own motion \citep{billah2020bioinspired}. Literature shows the STMD neurons found in insects are also highly responsive to a target's size, velocity, contrast, relative distance and direction of motion \citep{evans2022dragonfly}. The STMD neurons also respond to a single target at any particular time \citep{wiederman2013selective}. From the cascaded model in \citet{wiedermann2013biologically}, it is evident that the neural pathways for both optic flow and STMDs can be correlated, or even be assumed to be cascaded one after the other in an EMD-ESTMD formation. STMD neurons process the optic flow patterns to extract relevant information about the surroundings. This control-oriented study uses a concise model of STMD neurons' output, in particular as mathematical maxima ($\max$) and argument of maxima ($\argmax$) operators on the optic flow signal from its preceding neural pathway \citep{billah2022modeling}. The STMD-model thus identifies the maximum magnitude of optic flow signal and its corresponding angular position from the entire visual field of view as
\begin{align} \label{e:stmd_model}
    y_\mathrm{stmd,1} &= \max_{\gamma,\beta}(\dot{\vec{Q}}) \nonumber \\
    y_\mathrm{stmd,2} &= \argmax_{\gamma,\beta}(\dot{\vec{Q}}),
\end{align}
where $\dot{\vec{Q}}$ is the optic flow signal defined in a continuous, body-frame-relative spherical coordinate system as developed in \citet{Humbert2005}, and $\gamma$ and $\beta$ are the respective azimuth and elevation angles (Fig.~\ref{f:azimuthElevation}). This proposed STMD-model captures the STMD neurons' dependency on target's size, velocity, relative distance and direction of motion, along with the single target selection property of the STMD neurons. 

For initial analysis of the implications of Eqn.~\eqref{e:stmd_model} in a multi-agent system, the STMD-model is implemented in a group of homogeneous agents (in simulation and ground robots experiments) constrained to horizontal planar motion and the STMD neurons are assumed to be perfect sensors with no temporal or frequency dependence, resulting in zero time delay and signal amplification over a wide range of frequencies. Boundary walls or obstacles are not considered. 

For horizontal planar motion, $\gamma$ spans between $[-\pi,\pi]$, $\beta=0$ and the planar optic flow equation for multi-agent environment can be expressed as \citep{billah2020bioinspired}
\begin{align} \label{e:opticflow}
\dot{\vec{Q}}(\gamma,\vec{q}_{a},&\dot{\vec{q}}_a,\vec{q}_{b},\dot{\vec{q}}_b)=-\dot{\theta}_{a} - \mu(\gamma,\vec{q}_{a},\vec{q}_{b}) \dot{\theta}_{b}
\nonumber \\							 +&\mu(\gamma,\vec{q}_{a},\vec{q}_{b})\{(\dot{x}_{b}-\dot{x}_{a})\sin\gamma - (\dot{y}_{b}-\dot{y}_{a})\cos\gamma)\},
\end{align}
where ($\vec{q}_a,\dot{\vec{q}}_a$) and ($\vec{q}_b,\dot{\vec{q}}_b$) are the states of the viewer and the target agents respectively, such that $\vec{q}_{a}=[x_a,y_a,\theta_a]$ and $\vec{q}_{b}=[x_b,y_b,\theta_b]$. The nearness function $\mu$ is the inverse of the magnitude of the vector $\vec{r}_{ab}$ connecting the viewer and the target agents
\begin{align} \label{e:muGeneral}
    \mu(\gamma,\vec{q}_{a},\vec{q}_{b}) = \frac{1}{|\vec{r}_{ab}|}.
\end{align}

\subsection{Bio-inspired Multi-Agent Feedback Control System} \label{ssec:feedback_law}
The STMD-model detects and outputs a selected target agent in terms of the maximum optic flow magnitude ($y_\mathrm{stmd,1}$) and its corresponding angular position ($y_\mathrm{stmd,2}$). A feedback law to close the loop must then specify how the STMD-output is used to regulate the agent's motion. Dragonflies use STMD neurons during flight to pursue and encounter smaller prey and conspecifics \citep{evans2022dragonfly} and a feedback guidance law from insect pursuer-evader models known as `pure pursuit' strategy in literature \citep{varennes2020two} was chosen to define the higher-order neural feedback following the STMD-output. The feedback rule is given by
\begin{align}\label{e:controlLaw}
    u_i=K[{y}_\mathrm{stmd,2}]_i = K [\argmax_{\gamma,\beta}(\dot{\vec{Q}_i})] ,
\end{align}
where $u_i$ is the feedback law for $i^{th}$ subsystem, $i=1,2,\ldots,N$, and $K>0$ is an arbitrary scalar feedback gain. This proportional feedback rule structure aims to maintain a constant bearing angle between the target and the pursuer.    

\begin{figure}[ht] \centering \includegraphics[width=\linewidth]{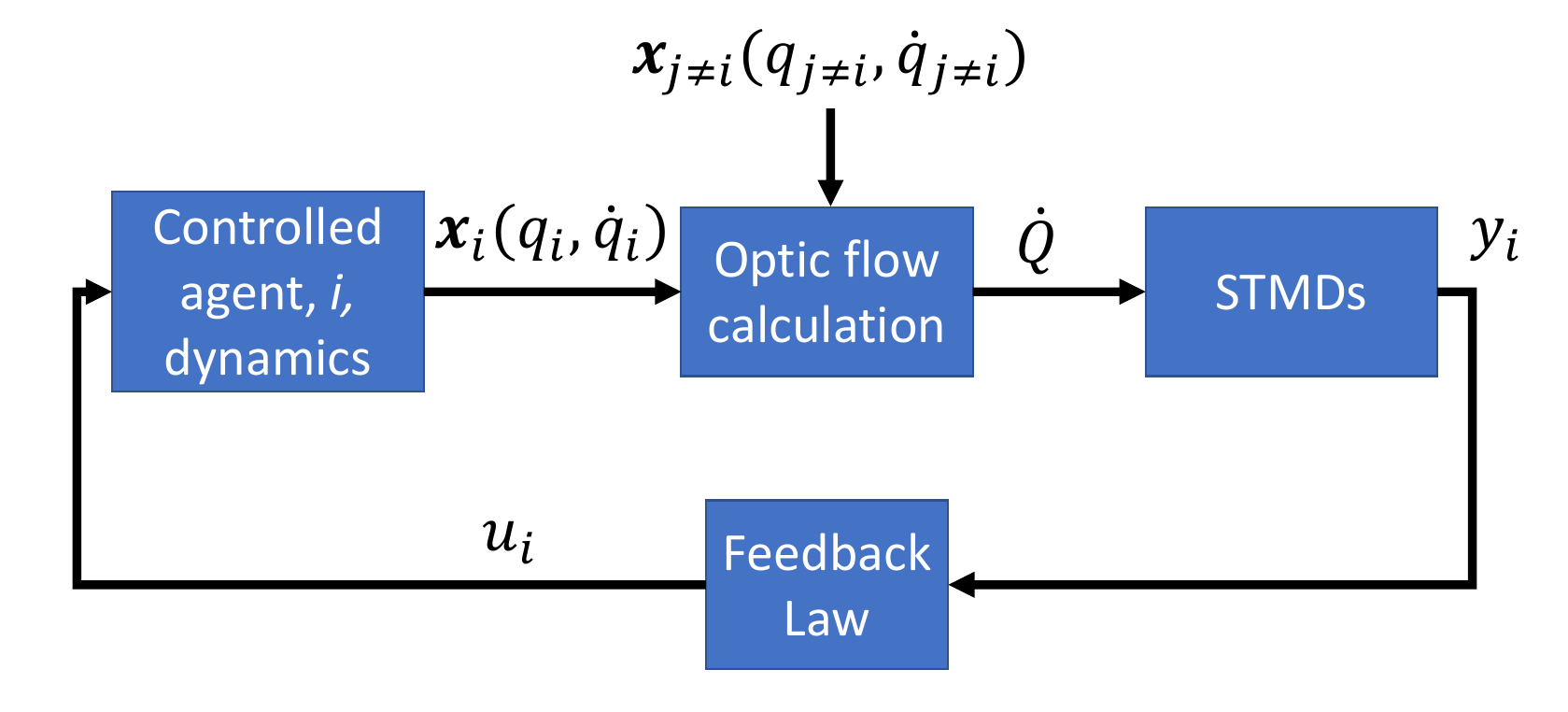}\caption{Small-target motion reactive (STMR), a bio-inspired multi-agent closed-loop feedback model based on STMD neurons as sensory mechanism.}\label{f:system_model} \end{figure}

The selection of one `target' agent based on the STMD-model output results in a continuous-time switching system, where the selection of target by a single agent switches between different neighboring agents over time depending on the highest magnitude of optic flow generated by the neighboring agents. The resulting bio-inspired multi-agent feedback control system is illustrated in Fig.~\ref{f:system_model}.   

\subsection{Stability Analysis of the Multi-Agent System} \label{sec:stability_proof}

\subsubsection{Bi-agent stability} \label{ssec:stability_bi_agent}
For stability analysis of the multi-agent feedback system, we consider differential-drive ground vehicles set at a constant forward speed $v$, following the non-holonomic constraint, whose kinematics can be defined for 2D horizontal planar motion as
\begin{align} \label{eq:groundVehicleDynamics}
    \dot{x_i} &= v_i \cos\theta_i \nonumber\\ 
    \dot{y_i} &= v_i \sin\theta_i \nonumber\\
    \dot{v_i} &= 0 \nonumber\\
    \dot{\theta_i} &= u_i,
\end{align}
where $u_i$ is the feedback control input given in Eqn.~\ref{e:controlLaw} and $v_i>0$ is the fixed forward speed.

\begin{figure}[htbp]
    \centering \includegraphics[width=\linewidth]{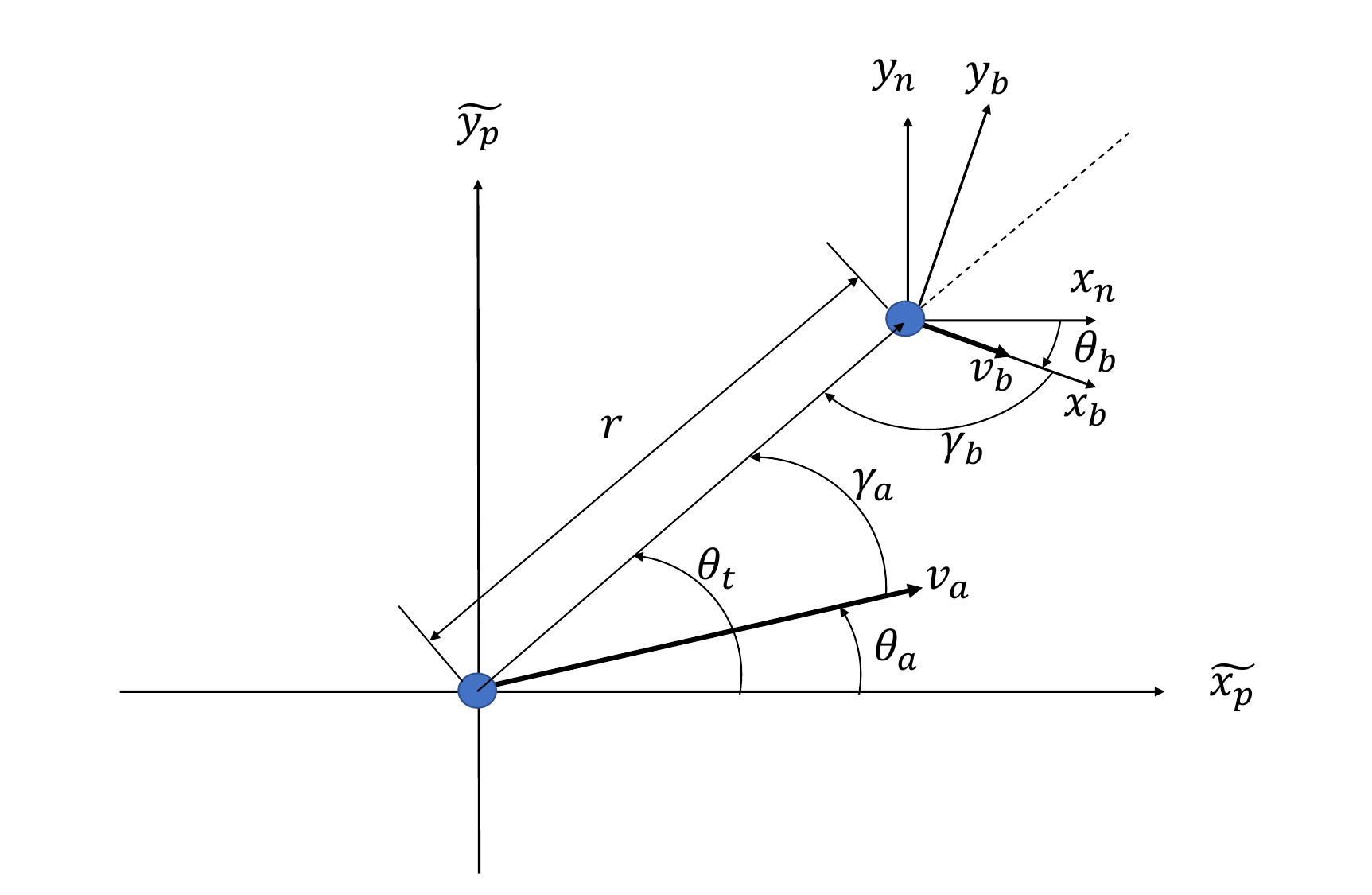}
    \caption{Geometry of the bi-agent problem.} \label{fig:geometry}
\end{figure}

First consider a bi-agent system with only two agents interacting with each other. Let the position error, defined as the distance between the agents, be given as
\begin{align}
    \Tilde{x} = x_a - x_b, \nonumber \\ 
    \Tilde{y} = y_a - y_b.
\end{align}

Let $\theta$ be the angle of the line-of-sight (LOS) of agent b w.r.t. agent a, then from Fig.~\ref{fig:geometry}
\begin{align}
    \Tilde{x} &= r \cos \theta_t, \nonumber \\ 
    \Tilde{y} &= r \sin \theta_t.
\end{align}

Differentiating error $\Tilde{x}$ and $\Tilde{y}$ with respect to time yields
\begin{align}
    \dot{\Tilde{x}} &= \dot{r} \cos \theta_t - r \sin \theta_t.\dot{\theta}_t = v_a \cos \theta_a - \dot{x}_b, \nonumber \\ 
    \dot{\Tilde{y}} &= \dot{r} \sin \theta_t + r \cos \theta_t.\dot{\theta}_t = v_a \sin \theta_a - \dot{y}_b.
\end{align}

Rewriting the equations for $\dot{r}$ and $\dot{\theta}$ yields
\begin{align} \label{e:two}
    \dot{r} &= v_a \cos (\theta_a - \theta_t) - (\dot{x}_b \cos(\theta_t) + \dot{y}_b \sin{\theta_t}), \nonumber \\ 
    \dot{\theta} &= \frac{1}{r} (v_a \sin (\theta_a - \theta_t) + (\dot{x}_b \sin(\theta_t) - \dot{y}_b \cos{\theta_t})).
\end{align}

Substituting $\dot{x}_b = v_b \cos \theta_b$ and $\dot{y}_b = v_b \sin \theta_b$ in Eqn.~\eqref{e:two} simplifies to  
\begin{align} \label{e:three}
    \dot{r} &= v_a \cos (\theta_a - \theta_t) - v_b \cos(\theta_t-\theta_b), \nonumber \\ 
    \dot{\theta} &= \frac{1}{r} (v_a \sin (\theta_a - \theta_t) + v_b \sin(\theta_t-\theta_b)).
\end{align}

The feedback law for horizontal planar motion is given by 
\begin{align} \label{e:feedback}
    \dot{\theta}_i = K[{y}_\mathrm{stmd,2}]_i = K_i\gamma_i,
\end{align}
where $K$ is the feedback gain constant and $\gamma$ is the azimuth angle. Elevation angle, $\beta=0$ for horizontal planar motion.  

From Fig.~\ref{fig:geometry}, the azimuth angles $\gamma_a$ and $\gamma_b$ can be defined as
\begin{align} \label{e:gamma}
    \gamma_a &= \theta_t - \theta_a, \nonumber \\ 
    \gamma_b &= \pi + \theta_b - \theta_t.
\end{align}

Since both the agents have the same forward speed, $v_a=v_b=v$. Let the nearness function be defined as $\mu=1/r$. For notational convenience, define $\alpha= v\mu$.

Using Eqns.~\eqref{e:three} and \eqref{e:feedback}, and differentiating Eqn.~\eqref{e:gamma}, the error dynamics with state variables $\gamma_a$ and $\gamma_b$ can be expressed in the polar coordinate as
\begin{align}
    \dot{\gamma_a} &= \alpha \sin \gamma_a + \alpha \sin \gamma_b - K_a \gamma_a, \nonumber \\ 
    \dot{\gamma_b} &=  K_b \gamma_b - \alpha \sin \gamma_a - \alpha \sin \gamma_b. 
\end{align}

Assume only one input to the system and let $K_b=0$, making the motion of point B a straight line. Then, linearizing the above system about the origin gives 
\begin{align}
    \begin{bmatrix} \dot{\gamma}_a \\ \dot{\gamma}_b \end{bmatrix} = 
    \begin{bmatrix} \alpha - K_a & \alpha \\ -\alpha & -\alpha \end{bmatrix} 
    \begin{bmatrix} \gamma_a \\ \gamma_b   \end{bmatrix}.
\end{align}
From the characteristic equation of this system, it can easily be seen that the origin is exponentially stable for $K_a>0$. 

\subsubsection{Multi-agent stability} \label{ssec:stability_multi-agent}
Now, to analyze the multi-agent switching system with $N$ agents, \textit{Theorem 2} from \citet{veer2019switched} is considered which describes stability of switched systems with multiple equilibria under disturbances. The theorem predicts bounded trajectories for input-to-state stable (ISS) subsystems under average dwell time bounds. The average dwell time $N_a$ is defined in terms of the number of switches $N_{\sigma} (t,\underbar{t}) \in \mathbb{Z}_+$ over the interval $[\underbar{t},t) \subset \mathbb{R}_+$ as
\begin{align} \label{e:average_dwell_time}
    N_{\sigma}(t,\underbar{t}) \leq N_0 + \frac{t-\underbar{t}}{N_a} \qquad \forall t\geq \underbar{t} \geq 0,
\end{align}
where $N_0>0$ is a finite constant.

The bounds for average dwell time is given by
\begin{align} \label{e:dwell_time_constr}
    N_0 \geq 1 \quad \mbox{and} \quad N_a \geq \Bar{N_a},
\end{align}
where $\Bar{N_a}>0$ represents the lower bound on the average dwell time of the system. For our system, we define $\Bar{N_a}$ for any constant $\epsilon$ in the open interval $(0,\lambda)$ as
\begin{align*}
    \Bar{N_a} = \frac{\ln\mu(k)}{\lambda-\epsilon},
\end{align*}
where $\lambda>0$ and $\mu(k)\geq 1$ are defined in \citet{veer2019switched}. 

\subsection{Simulation Study}\label{ssec:simulation}
Simulation of the multi-agent switched system in Fig.~\ref{f:system_model} under average dwell time bound from Eqn.~\eqref{e:dwell_time_constr} is implemented in Matlab. A total of $N$ homogeneous agents are considered in the 2D planar motion simulation environment and assumed to be following the kinematics of ground vehicles described in Eqn.~\eqref{eq:groundVehicleDynamics}. For each agent, this framework results in $N-1$ identical subsystems that the STMD output can switch among. The simplified simulation assumes sparse environment with no background or obstacles, and no explicit collision avoidance algorithm in place (outside of any given by the feedback law). All of the agents are set at the same fixed forward speed $v_i$ with arbitrary initial positions and orientations. Measurement of optic flow and STMD output uses the idealized equations from Eqn.~\eqref{e:opticflow} and Eqn.~\eqref{e:stmd_model} respectively. Target switching for each agent is allowed only if the average dwell time bound is satisfied.

\subsection{Experimental Study} \label{ssec:experiment}

\begin{figure}[htbp] \centering 
	\begin{subfigure}[h]{0.5\linewidth} \includegraphics[width=\textwidth]{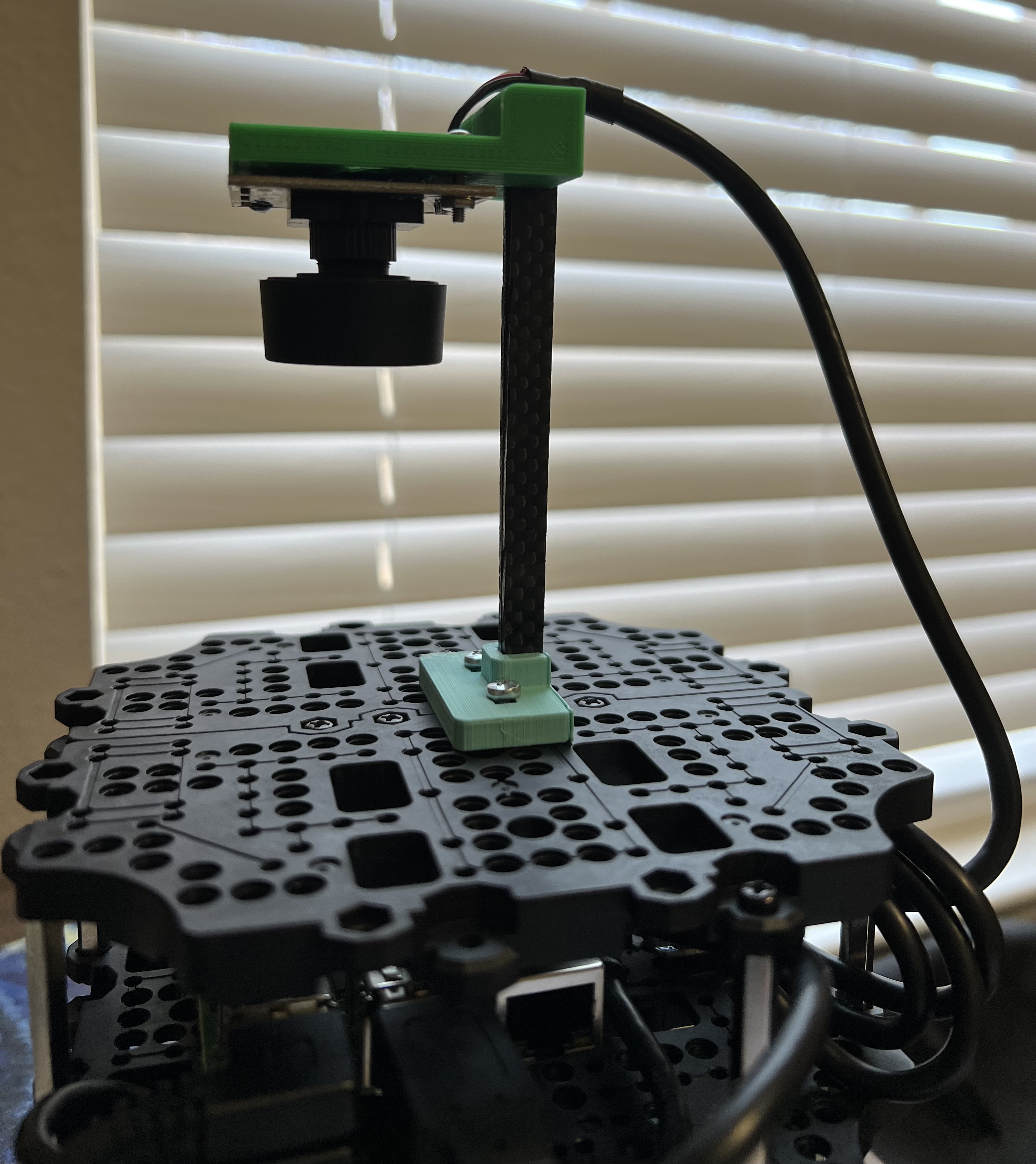} 
    \end{subfigure} 
	\begin{subfigure}[h]{0.9\linewidth} \includegraphics[width=\textwidth]{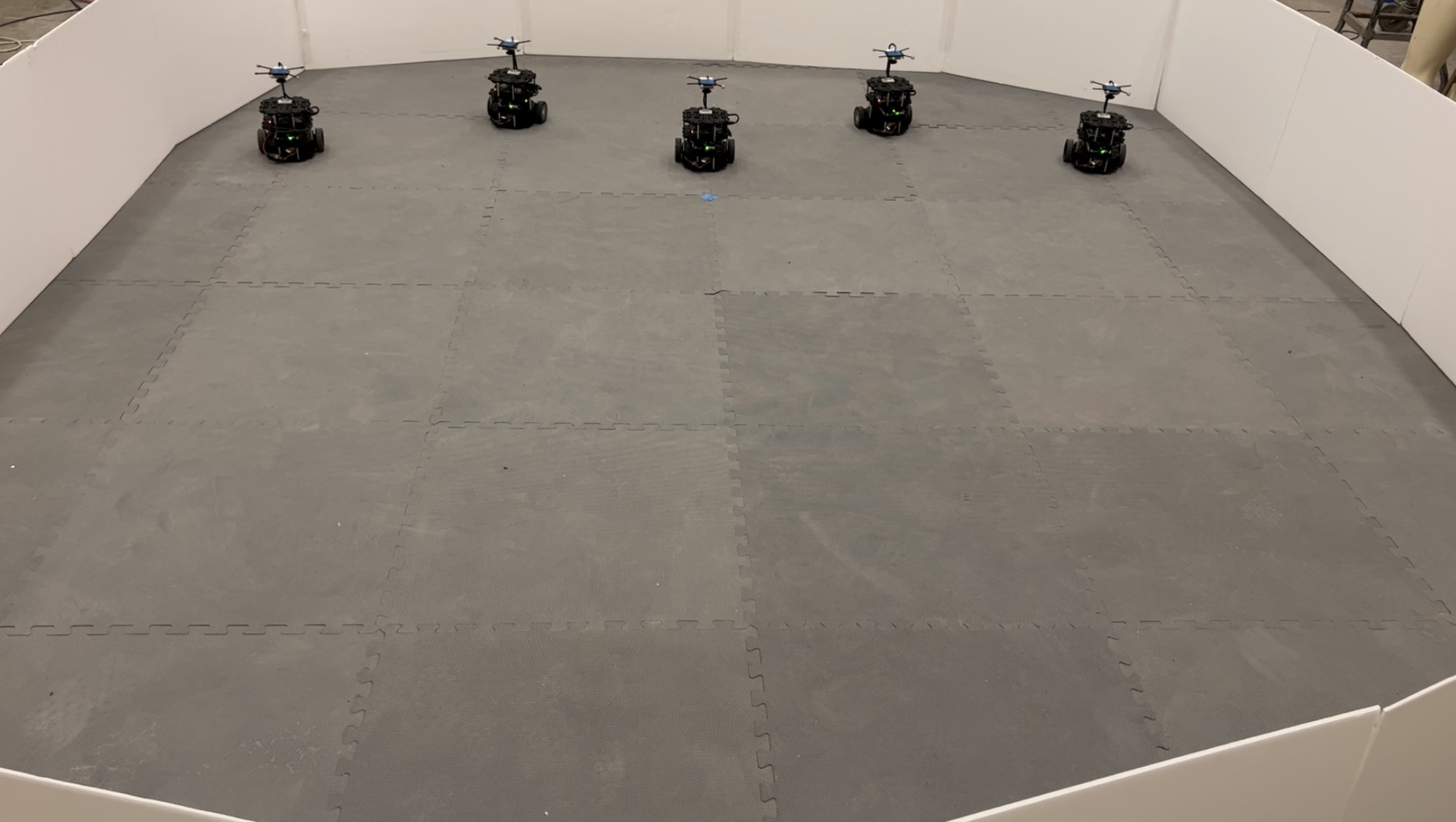}
    \end{subfigure}
    \caption{Turtlebot (top), the ground robot used for implementation of the feedback system in Fig.~\ref{f:system_model}, equipped with an inverted $360^{\circ}$ camera for optic flow measurement, and experimental setup (bottom) showing five Turtlebots in the experiment arena.} \label{f:exp_setup}
\end{figure}

For real-time experimental implementation on ground robots, five Turtlebots \citep{amsters2020turtlebot} equipped with panoramic cameras and running the decentralized feedback controller onboard are used (Fig.~\ref{f:exp_setup}, top), starting from arbitrary initial conditions that were kept constant for all the experiments. The experiment arena uses white boundary walls for low visual texture to ensure the swarm agents' motions are the primary sources of optic flow (Fig.~\ref{f:exp_setup}, bottom). The Turtlebots are operated by Raspberry Pi 3b running Raspbian OS. Python, C++, and ROS Kinetic are used to implement the onboard feedback system processes, including onboard image acquisition and processing, feedback control calculation and implementation. The agents have a maximum linear velocity of 22 cm/s and maximum rotational velocity of 2.84 rad/s. No explicit collision avoidance algorithm is implemented on the Turtlebots. The experimental arena spans 8~x~12 ft. The camera used is Picam360 with a vertical FOV of $220^{\circ}$ and a horizontal FOV of $360^{\circ}$. It captures 640~x~480 pixel images at 20 fps. The raw, circular image is then distortion corrected into a panoramic image using image processing. The Picam360 is mounted face down to ensure the target agents occupied a large FOV in the dewarped images. A single horizontal line from the input image contains all the target agents and is used to perform onboard optic flow calculation. Optic flow calculation from the camera image input is performed using the Gunnar Farneb\"ack algorithm using the opencv\_apps ROS package. The Gunnar Farneb\"ack algorithm calculates dense optic flow at fixed points in the visual input based on pixel intensity changes between successive image frames. Refer to \citep{billah2023robustness} for more details of the experimental setup and optic flow measurement. 

A remote desktop running Linux-based Ubuntu 16.04 initializes all the agents at the start of the experiment. Each experimental trial consisted of 15 seconds.

Two cases have been explored in this setup for comparison: (a) pure pursuit STMR, which uses the feedback rule defined in Eqn.~\eqref{e:controlLaw}, and (b) motion camouflage STMR, which uses the feedback rule for motion camouflage defined in \citet{justh2006steering}. Both cases use STMD-based sensory model for optic flow processing, given in Eqn.~\ref{e:stmd_model}. The scalar feedback gains $K_i$ were chosen arbitrarily, and the average dwell time was chosen to be 1.5s, based on the stability analysis in Sec.~\ref{sec:stability_proof}. All the Turtlebots were set with a fixed forward speed of $v_i=11$ cm/s. Gains $K_i$ and average-dwell time were also the same for all agents. The initial conditions for all agents were chosen arbitrarily and kept constant for all trials of both the cases. 

\subsection{Comparison Study} \label{ssec:comp}
For comparison of the STMR framework with contemporary swarm approaches, three other multi-agent systems are considered: the Vicsek model \citep{Vicsek1995}, the Cucker-Smale (C-S) model \citep{Cucker2007} and the multi-agent optomotor response (MA-WFI) model developed previously by the authors \citep{billah2020bioinspired} that uses wide-field integration (WFI) of optic flow signals for regulating an agent's motion relative to its neighbors. The Vicsek model updates agent orientation based on the average orientation of agents in its neighborhood, defined by a radius of interaction. The C-S model assumes inter-agent interaction occurs between all the agents in the environment, regardless of their inter-agent distances, and the agents update velocity based on a weighted function of the average velocity of all the other agents. The Vicsek model achieves orientation-matching whereas C-S model does velocity-matching. The MA-WFI model is designed to be able to do both orientation-matching and velocity-matching for a controlled-agent in an established formation of other agents based on the choice of basis functions. 

Two different environments are considered for comparison of feedback systems. For the first case, the MA-WFI simulation environment is used, where only one agent is controlled among other agents in an established formation. The controlled agent has an initial orientation offset, while the rest starts with no offset. The simulation is run four times, once for each of the different feedback control systems. For the second case of comparison, all agents are being controlled by the four different control systems starting from the same initial conditions as in the first case. The C-S agents have varying initial velocity, the other models have the same fixed forward speed for all agents during the entire simulation. 

For further comparison analysis of the different frameworks, we study the connectivity of a system of 50 agents governed by the four different multi-agent systems mentioned earlier: STMR, MA-WFI, C-S and Vicsek. For this study, the agents are converted to nodes ($V$) of a weighted graph $\mathcal{G}=(V,E,W)$, where an edge ($E$) exists between a pair of nodes only if the two agents interact with each other. Edge weights ($W$) are determined based on the feedback rules that the agents use. Algebraic connectivity (also called the Fiedler eigenvalue) is defined as the second-smallest eigenvalue of the Laplacian matrix of $\mathcal{G}$ \citep{godsil2001algebraic} and is a measure of how well connected the agents are among each other in a multi-agent system. Three swarm analysis metrics (swarm polarisation \citep{armbruster2017elastic}, average swarm heading and heading variance) are analyzed to provide a systematic performance comparison across the four frameworks.

\section{Results and Discussion}\label{sec:results&discussion}

\subsection{Results} \label{ssec:results}

\subsubsection{Simulation results} \label{sssec:simulation_results}
\begin{figure*}[htbp] \centering
	\begin{subfigure}[h]{0.30\linewidth} \includegraphics[width=\textwidth]{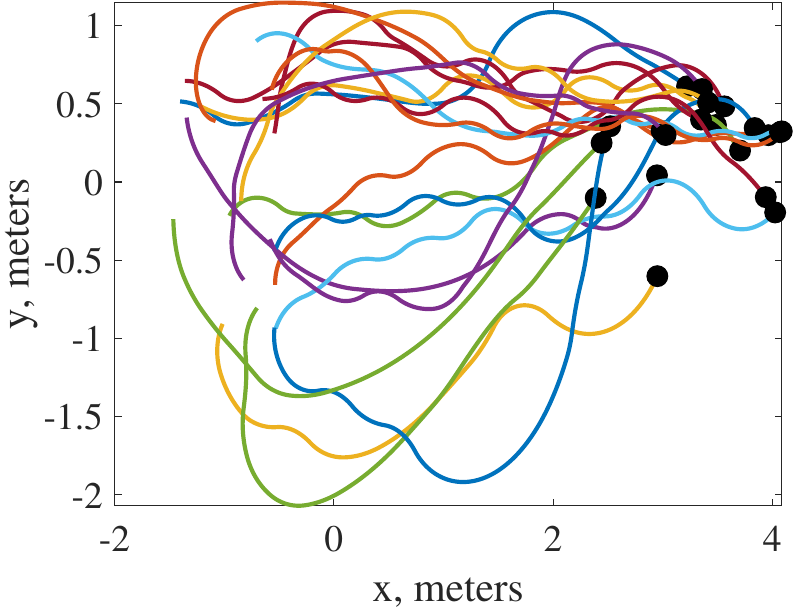} 
	\end{subfigure}
	\begin{subfigure}[h]{0.30\linewidth}
	\includegraphics[width=\textwidth]{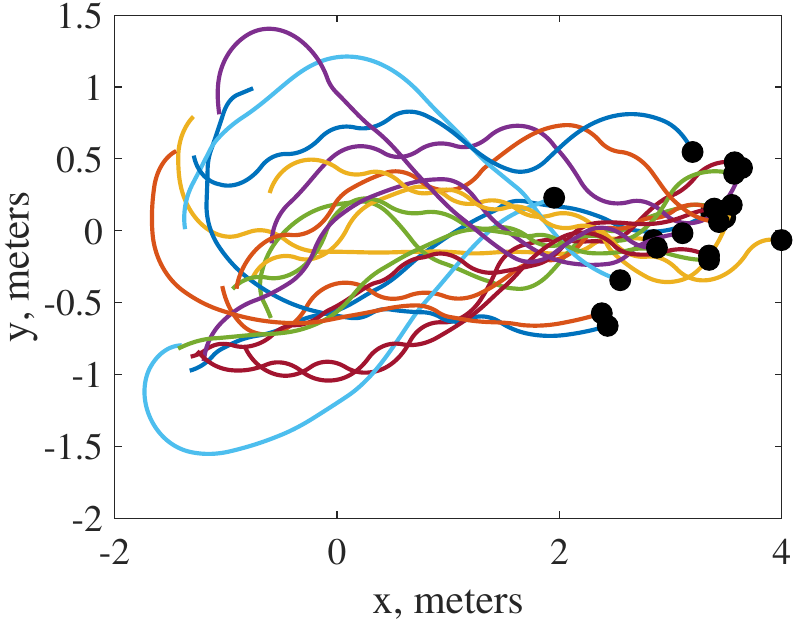} 
	\end{subfigure}
	\begin{subfigure}[h]{0.30\linewidth}
    \includegraphics[width=\textwidth]{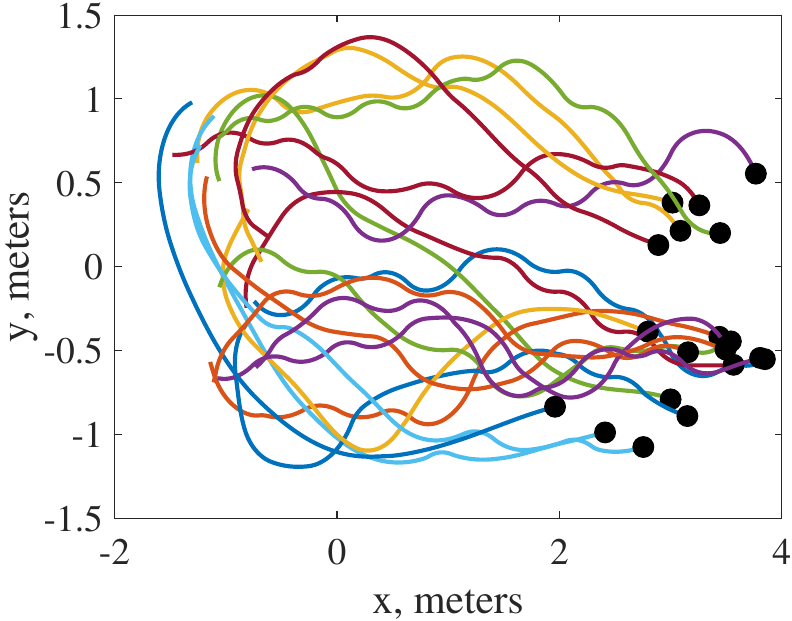}
	\end{subfigure}
	\begin{subfigure}[h]{0.30\linewidth} \includegraphics[width=\textwidth]{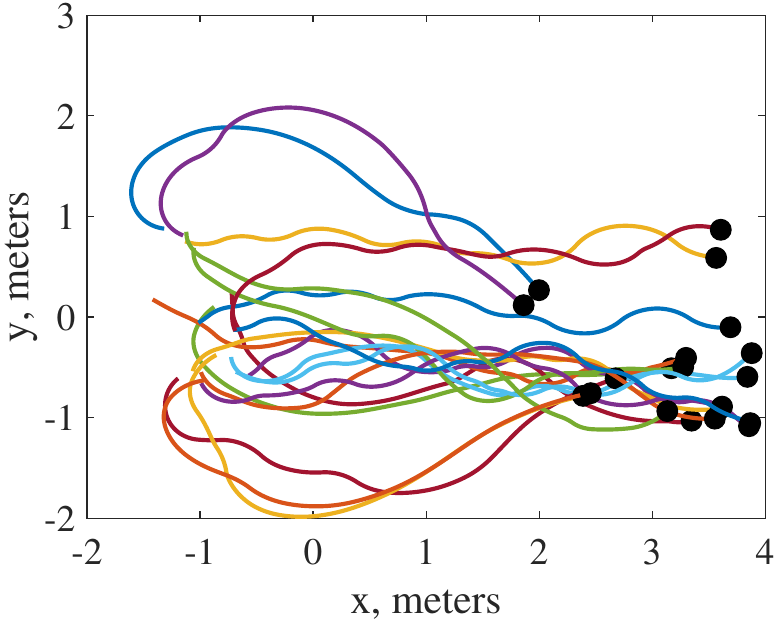} 
    \end{subfigure}
	\begin{subfigure}[h]{0.30\linewidth}
	\includegraphics[width=\textwidth]{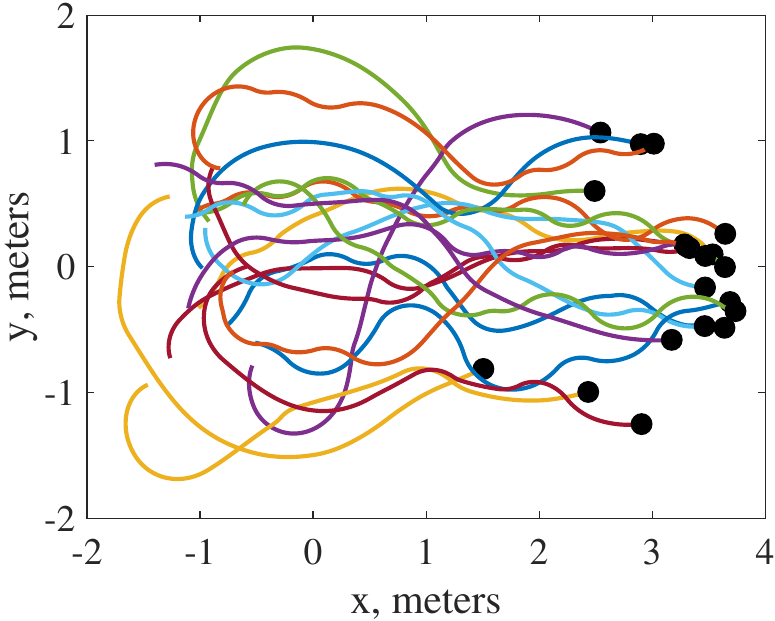} 
    \end{subfigure}
	\begin{subfigure}[h]{0.30\linewidth}
    \includegraphics[width=\textwidth]{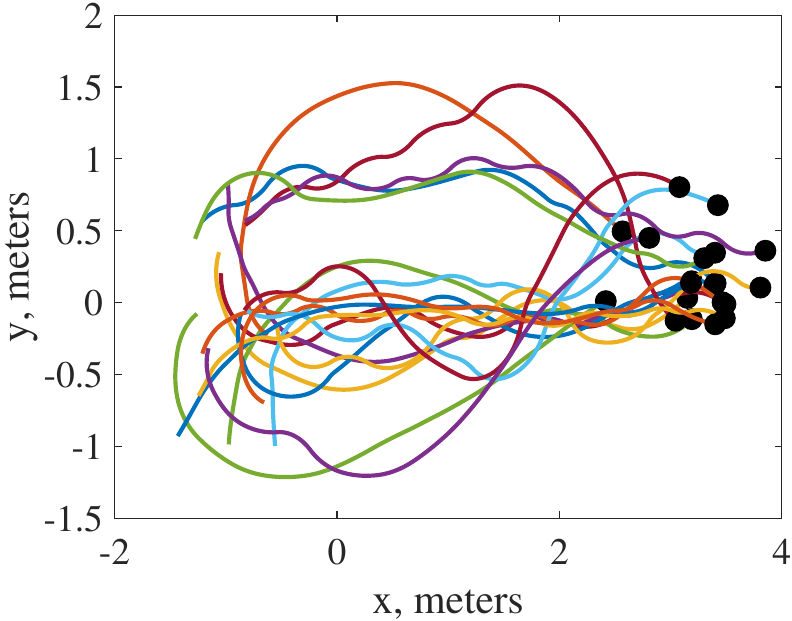}
	\end{subfigure}
\caption{Simulation results for the multi-agent switched system in Fig.~\ref{f:system_model} with the average dwell time constraints, showing trajectories of 20 agents for six different initial conditions.} 
\label{sim1a}
\end{figure*}

Simulation results of $N=20$ agents for six different initial conditions are presented in Fig.~\ref{sim1a}. Feedback gains and agents' forward speeds are set to $0.1$ and $10$ cm/s respectively. Parameters to determine the minimum average dwell time are set to: $\mu(k)=10$, $\lambda=1$ and $\epsilon=0.3$. Each simulation is run for 50 seconds.

\begin{figure*}[htbp] \centering
	\begin{subfigure}[h]{0.30\linewidth} \includegraphics[width=\textwidth]{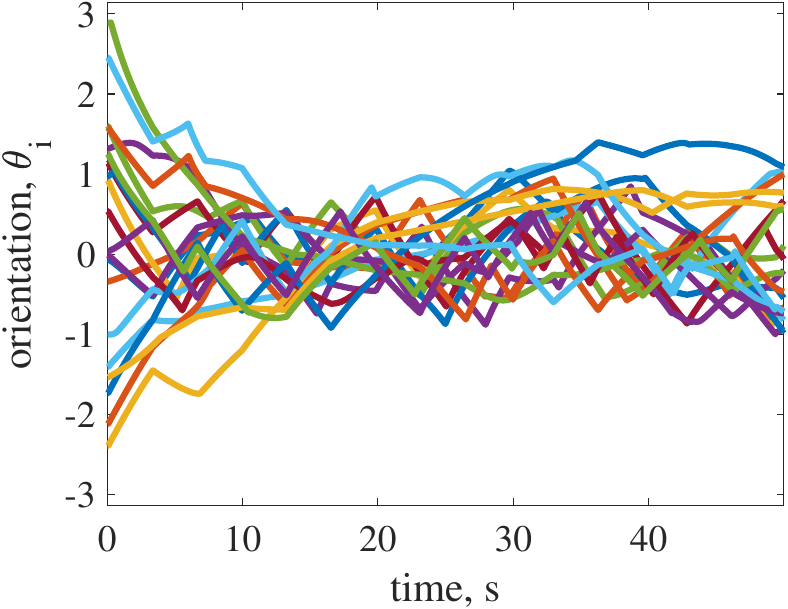} 
	\end{subfigure}
	\begin{subfigure}[h]{0.30\linewidth}
	\includegraphics[width=\textwidth]{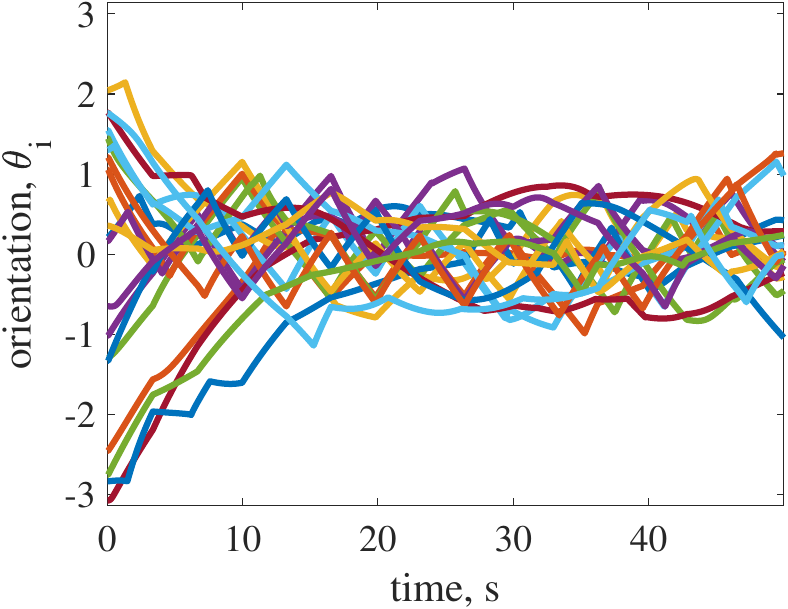} 
	\end{subfigure}
	\begin{subfigure}[h]{0.30\linewidth}
    \includegraphics[width=\textwidth]{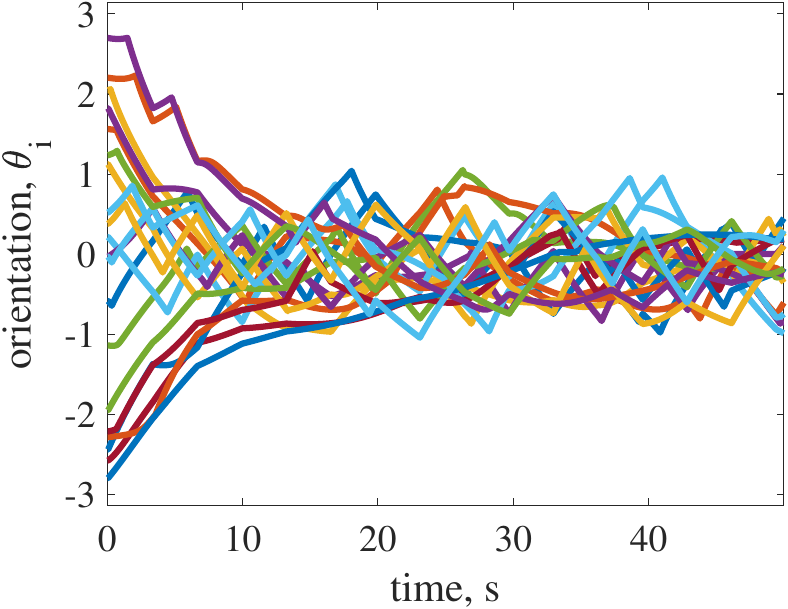}
	\end{subfigure}
	\begin{subfigure}[h]{0.30\linewidth} \includegraphics[width=\textwidth]{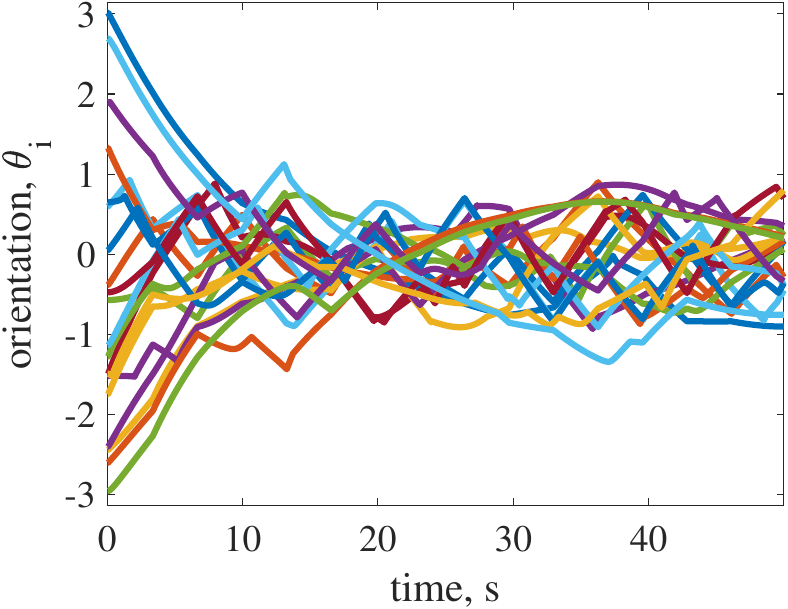} 
    \end{subfigure}
	\begin{subfigure}[h]{0.30\linewidth}
	\includegraphics[width=\textwidth]{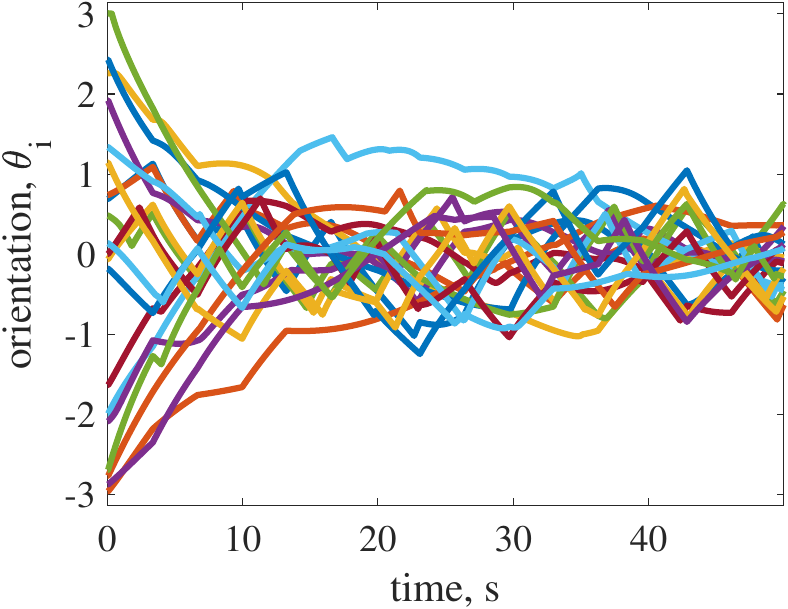} 
    \end{subfigure}
	\begin{subfigure}[h]{0.30\linewidth}
    \includegraphics[width=\textwidth]{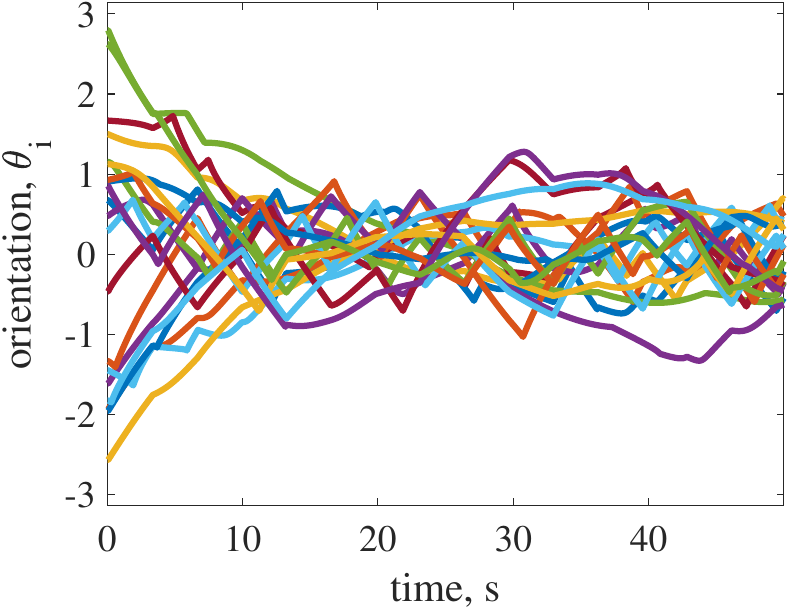}
	\end{subfigure}
\caption{Simulation results for the multi-agent switched system in Fig.~\ref{f:system_model} with the average dwell time constraints, showing orientation of 20 agents for the six initial conditions from Fig.~\ref{sim1a}.} 
\label{sim1b}
\end{figure*}

The markers in Fig.~\ref{sim1a} represent the final position of the agents while the solid lines represent their trajectories. Collision avoidance is not designed into the feedback rule, so collision between agents is observed in some instances. Fig.~\ref{sim1b} shows all the agents' orientations $\theta_i$ over the simulation period. The agents show general directional consensus and bounded $\theta_i$ for each of the different initial conditions.

\begin{figure}[htbp] \centering 
	\begin{subfigure}[h]{0.6\linewidth} \includegraphics[width=\textwidth]{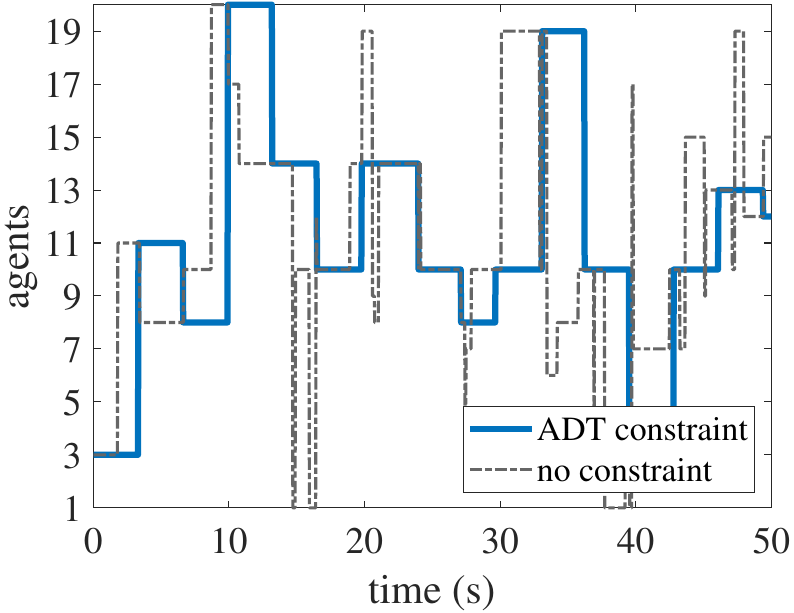} \end{subfigure} 
	\begin{subfigure}[h]{0.6\linewidth} \includegraphics[width=\textwidth]{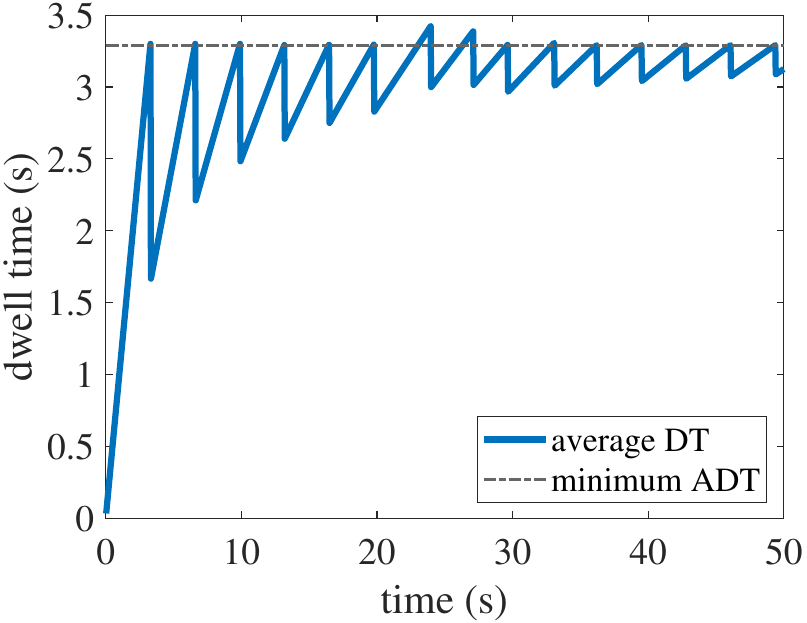} \end{subfigure}
    \caption{Simulation results. Switching signal plot (top) showing the agents ``tracked'' over time by a single agent (fifth agent from top in Fig.~\ref{sim1a} bottom center axes) in the multi-agent switched system with and without the dwell time constraints, and the average dwell time plot (bottom) for the same agent with the minimum average dwell time plotted.}    \label{f:simexp2}\end{figure}

Fig.~\ref{f:simexp2} (top) presents the switching signal plot with and without the average dwell time bounds. The plot indicates which agents are `tracked' over time by an arbitrary agent. Each agent `tracks' one other agent based on the STMD-output (Eqn.~\eqref{e:stmd_model}). Fig.~\ref{f:simexp2} (bottom) shows the average dwell time plot of an agent. The minimum average dwell time based on Eqn.~\eqref{e:dwell_time_constr} is also plotted. It is clear that switching between `tracked' agents occurs only when the average dwell time for that agent is above the minimum specified. After each switch, the average dwell time decreases as expected from Eqn.~\eqref{e:average_dwell_time}, resulting in the sawtooth-like shape of the plots. 

\begin{figure*}[htbp] \centering
	\begin{subfigure}[h]{0.30\linewidth} 
    \includegraphics[width=\textwidth]{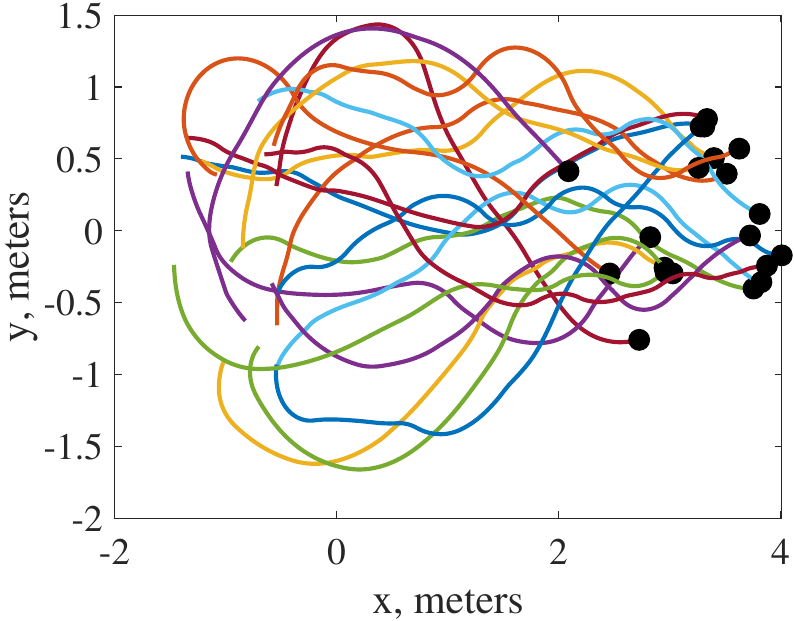} 
	\end{subfigure}
	\begin{subfigure}[h]{0.30\linewidth}
	\includegraphics[width=\textwidth]{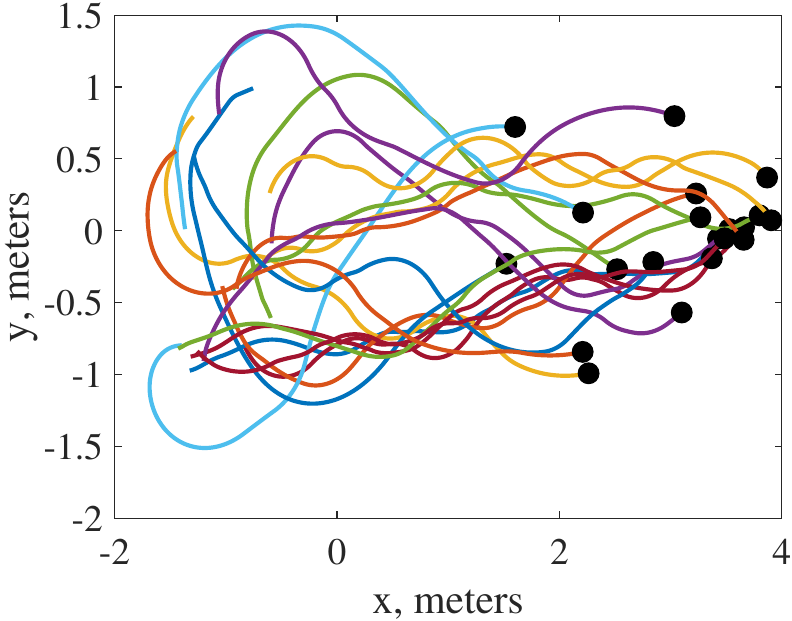} 
	\end{subfigure}
	\begin{subfigure}[h]{0.30\linewidth}
    \includegraphics[width=\textwidth]{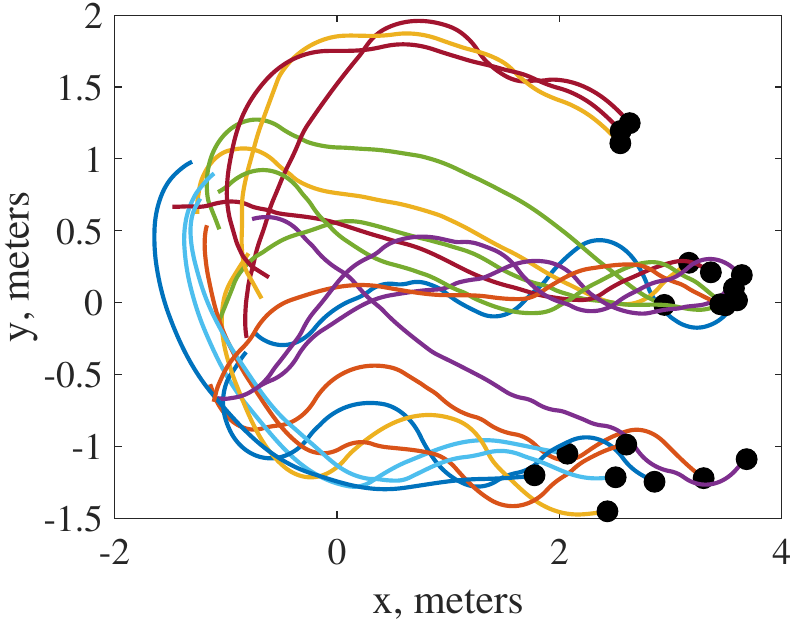}
	\end{subfigure}
	\begin{subfigure}[h]{0.30\linewidth} 
    \includegraphics[width=\textwidth]{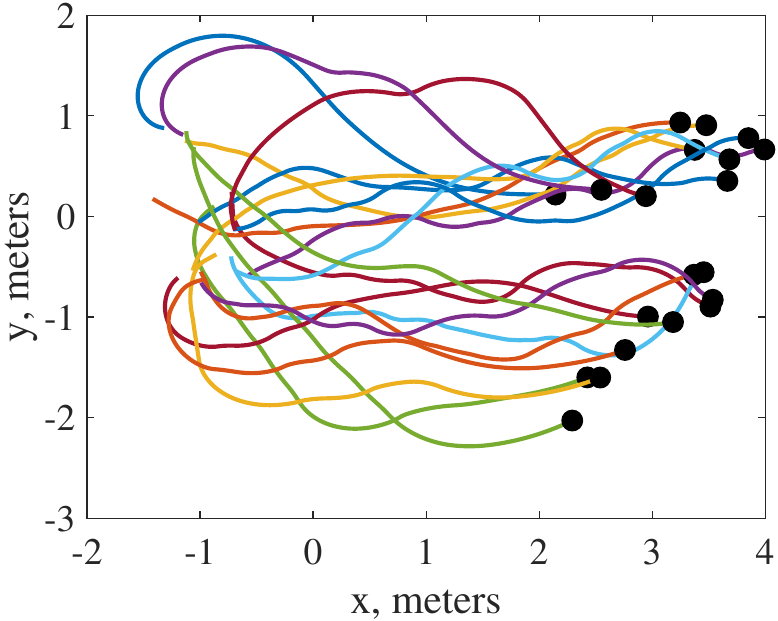} 
    \end{subfigure}
	\begin{subfigure}[h]{0.30\linewidth}
	\includegraphics[width=\textwidth]{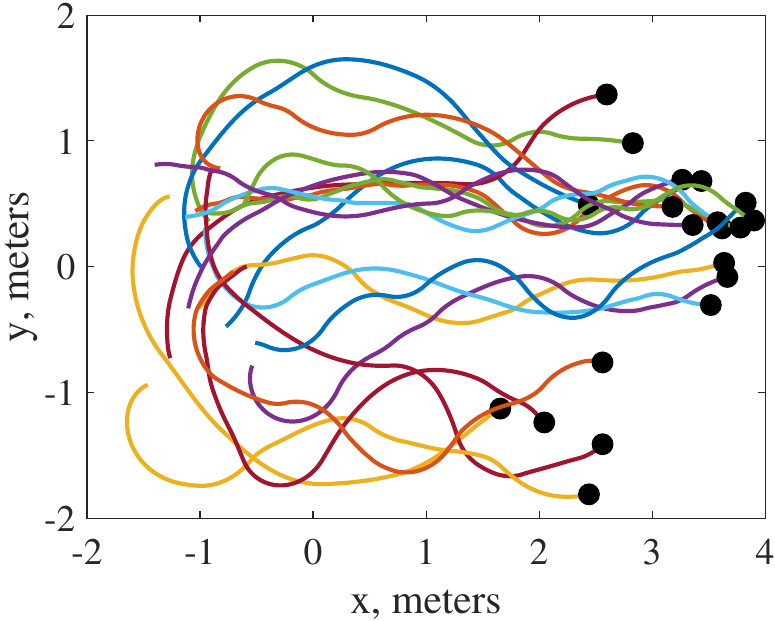} 
    \end{subfigure}
	\begin{subfigure}[h]{0.30\linewidth}
    \includegraphics[width=\textwidth]{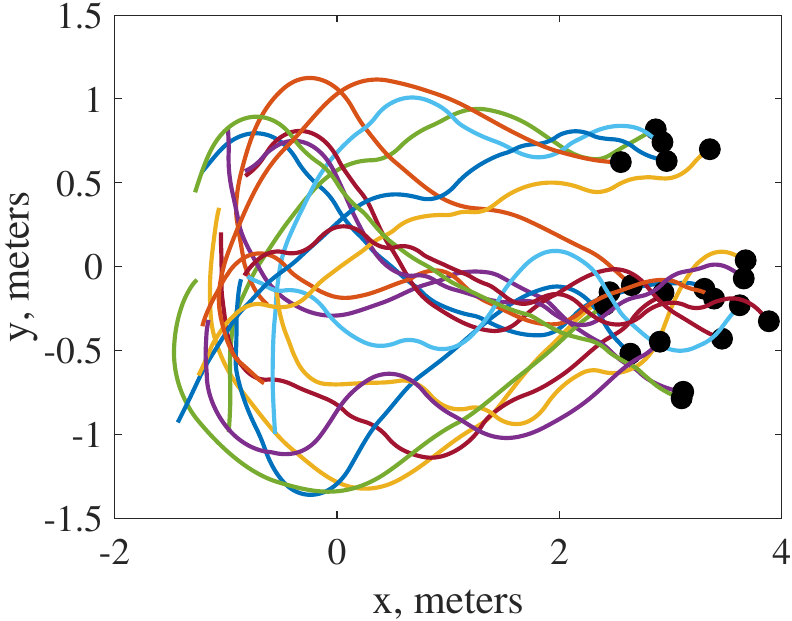}
	\end{subfigure}
\caption{Simulation results without the average dwell time constraints, showing trajectories of 20 agents for initial conditions from Fig.~\ref{sim1a}.} 
\label{sim1c}
\end{figure*}

Fig.~\ref{sim1c} shows the agent trajectories for simulation of the system without the average dwell time bounds, allowing the agents to switch instantaneously to whichever neighboring agent has the highest magnitude of optic flow. Initial conditions for the six cases were kept the same as those in Fig.~\ref{sim1a} for comparison. The agents still show general directional consensus over time, but one observes fragmentation into smaller groups of agents in some cases (Fig.~\ref{sim1c}, top right).

\subsubsection{Experimental results} \label{sssec:exp_results}
\begin{figure}[htbp] \centering 
	\begin{subfigure}[h]{0.65\linewidth} \includegraphics[width=\textwidth]{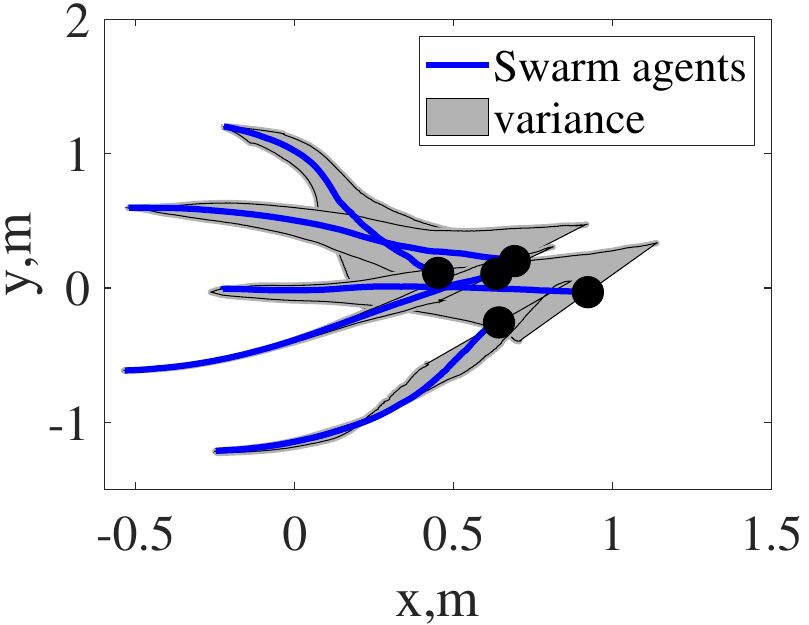}
    \end{subfigure} 
	\begin{subfigure}[h]{0.65\linewidth} \includegraphics[width=\textwidth]{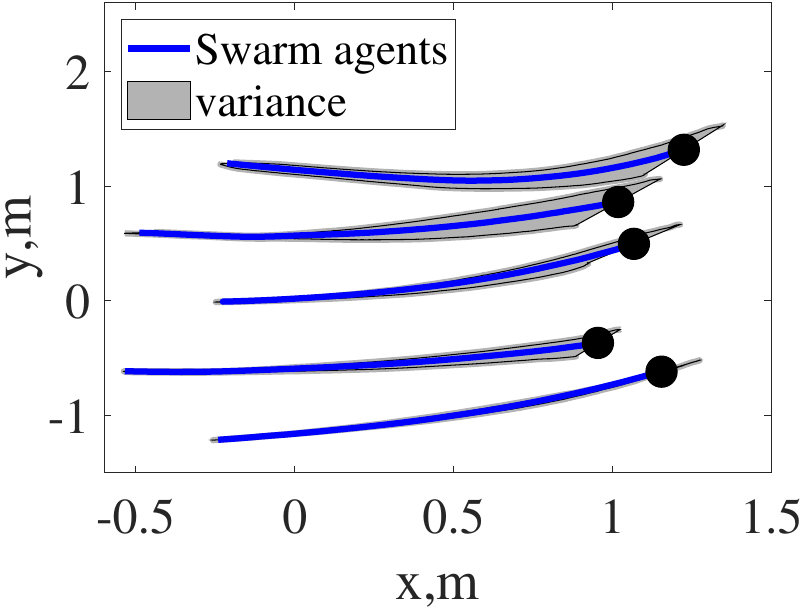}
    \end{subfigure}
    \caption{Experimental results for the implementation of multi-agent STMR feedback control system with two different feedback rules considered (15 trials each): (top) the pure pursuit feedback rule from Eqn.~\eqref{e:controlLaw} and (bottom) motion camouflage from \citep{justh2006steering}.}\label{f:STMD_exp}
\end{figure}

Experimental results for the two feedback rules are shown in Fig.~\ref{f:STMD_exp}. Blue lines indicate mean agent trajectories, and grey areas represent the variance over 15 trials for the same initial condition. The agents converge towards each other for pure pursuit STMR feedback in Fig.~\ref{f:STMD_exp} (top). Collision occurs for a minority of agents since no collision avoidance algorithm is being implemented. Unlike the simulation, collision has a greater effect in the experiments as ground vehicles continue to interfere after collision. Increasing initial agent separation and reducing the experiment duration to a relatively shorter time (15s) compared to the simulation (50s) allowed the multi-agent system to show convergence behaviors while avoiding artifacts from collision-induced behavior. Analyzing the effects and consequences of collision among agents is not within the scope of this study. 

For the case of using motion camouflage STMR feedback, results shown in Fig.~\ref{f:STMD_exp} (bottom), the agents do not converge on the same point or collide, and consistently demonstrate low variance in their trajectories across the 15 trials. The agents' group motion demonstrates higher polarisation and more consistent spatial separation.   

\subsubsection{Comparison study results} \label{sssec:comp_results}
\begin{figure} 
\centering\includegraphics[width=0.6\columnwidth]{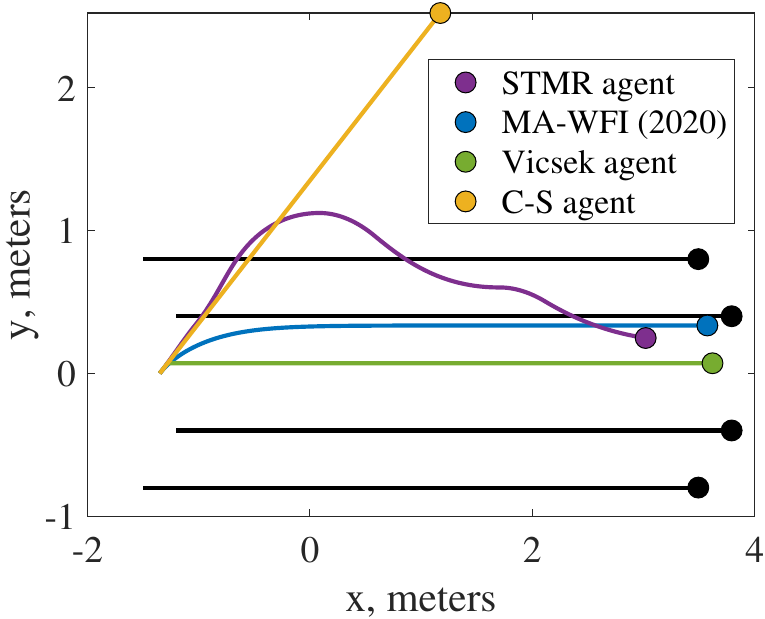} \caption{A single feedback-controlled agent surrounded by non-reactive neighbors with established formation; comparison of STMR feedback control system with the optomotor feedback control system from \citep{billah2020bioinspired}, the Vicsek model \citep{Vicsek1995} and the Cucker-Smale model \citep{Cucker2007}.} \label{f:comp_a} 
\end{figure}

\begin{figure}[htbp] \centering 
	\begin{subfigure}[h]{0.6\linewidth} \includegraphics[width=\textwidth]{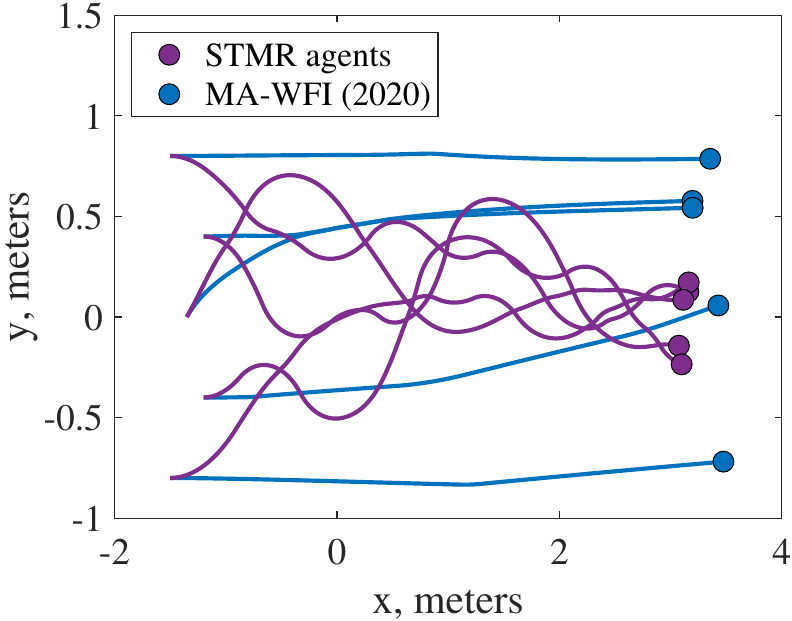}
    \label{f:comp_b}\end{subfigure}
    \begin{subfigure}[h]{0.6\linewidth} \includegraphics[width=\textwidth]{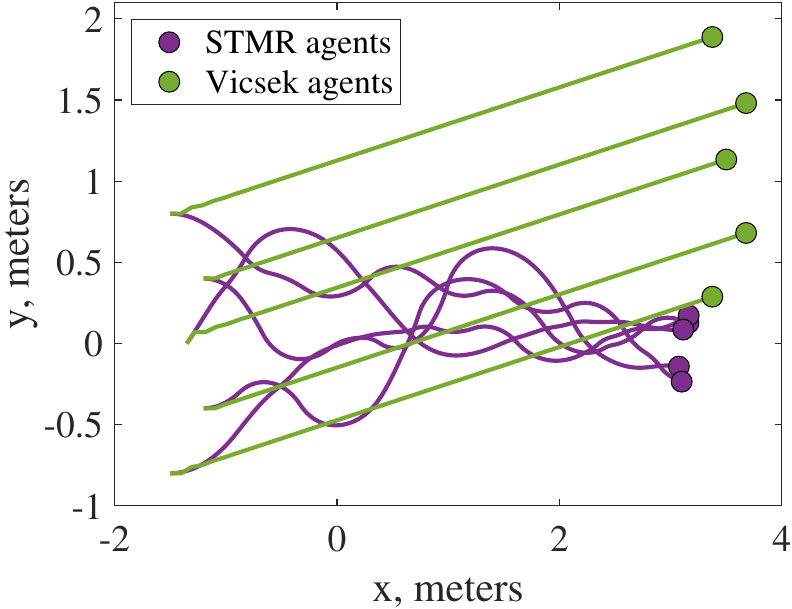}
    \label{f:comp_c}\end{subfigure}
    \begin{subfigure}[h]{0.6\linewidth} \includegraphics[width=\textwidth]{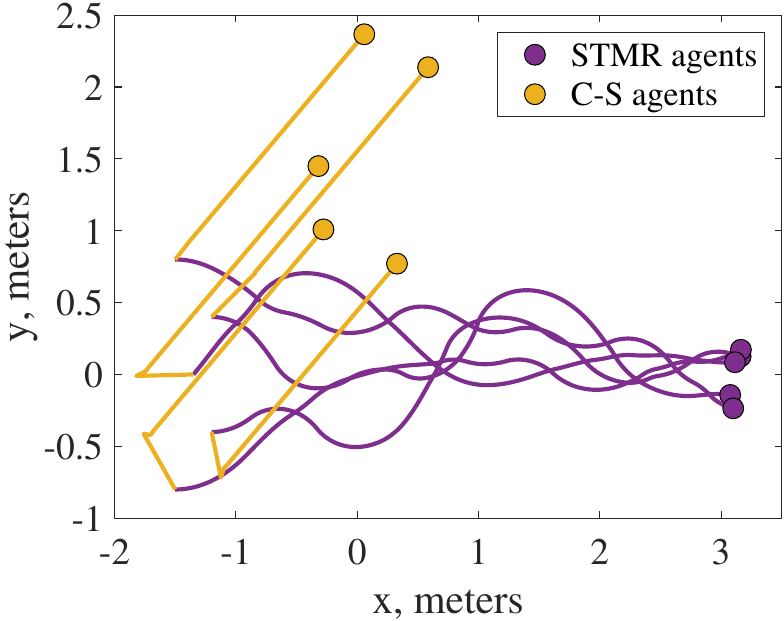}
    \label{f:comp_d}\end{subfigure}
    \caption{A collection of feedback-controlled agents all reactive to each other; comparison of the STMR feedback control system with the optomotor feedback control system from \citep{billah2020bioinspired}, the Vicsek model \citep{Vicsek1995} and the Cucker-Smale model \citep{Cucker2007}.}
    \label{f:compare}
\end{figure}

\begin{figure}[htbp] \centering 
	\begin{subfigure}[h]{0.6\linewidth} \includegraphics[width=\textwidth]{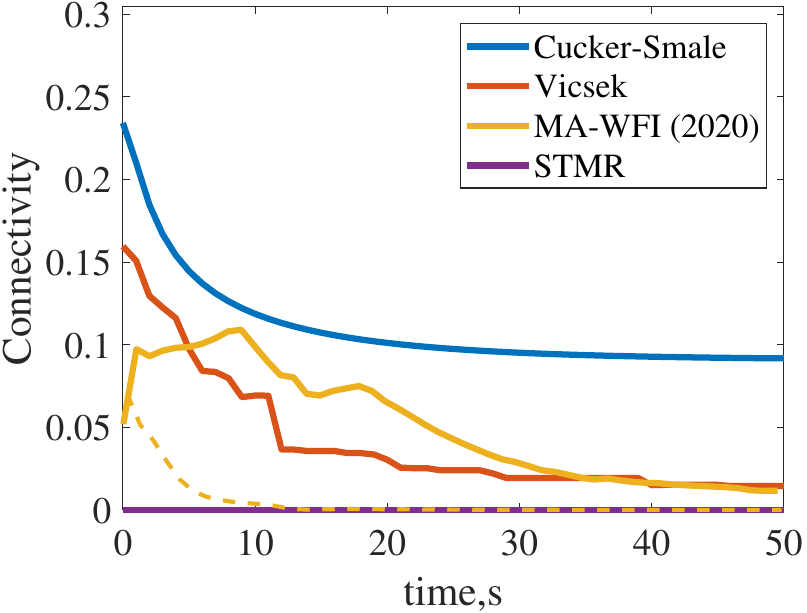}
    \caption{Connectivity of the underlying undirected graph.}\label{f:conn_comp_a}\end{subfigure} 
	\begin{subfigure}[h]{0.6\linewidth} \includegraphics[width=\textwidth]{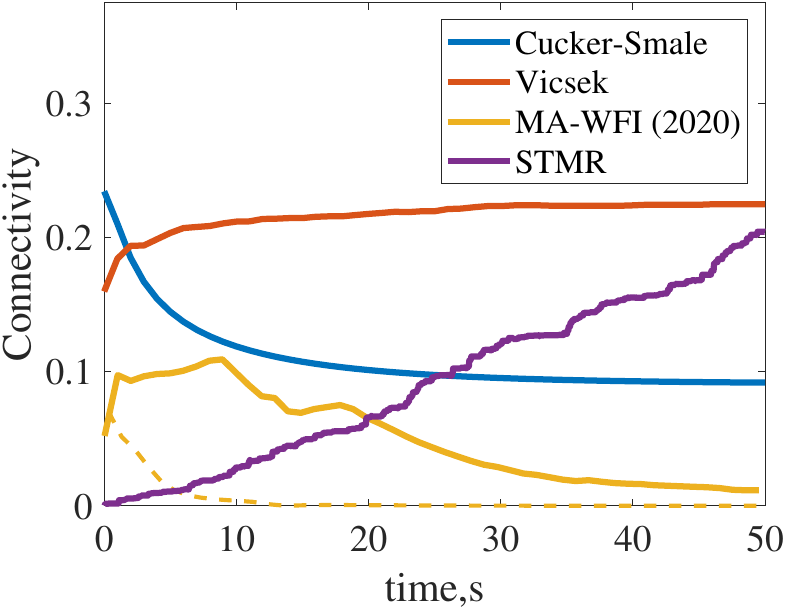}
    \caption{Connectivity of the union of underlying graphs.}\label{f:conn_comp_b}\end{subfigure}
    \caption{Comparison of algebraic connectivity (Fiedler eigenvalue) among the STMR feedback control system proposed in this paper, the optomotor feedback control system from \citep{billah2020bioinspired}, the Vicsek model \citep{Vicsek1995} and Cucker-Smale model \citep{Cucker2007}. }
    \label{f:connectivity_compare}
\end{figure}

\begin{figure*}[htbp] \centering 
	\begin{subfigure}[h]{0.32\linewidth} \includegraphics[width=\textwidth]{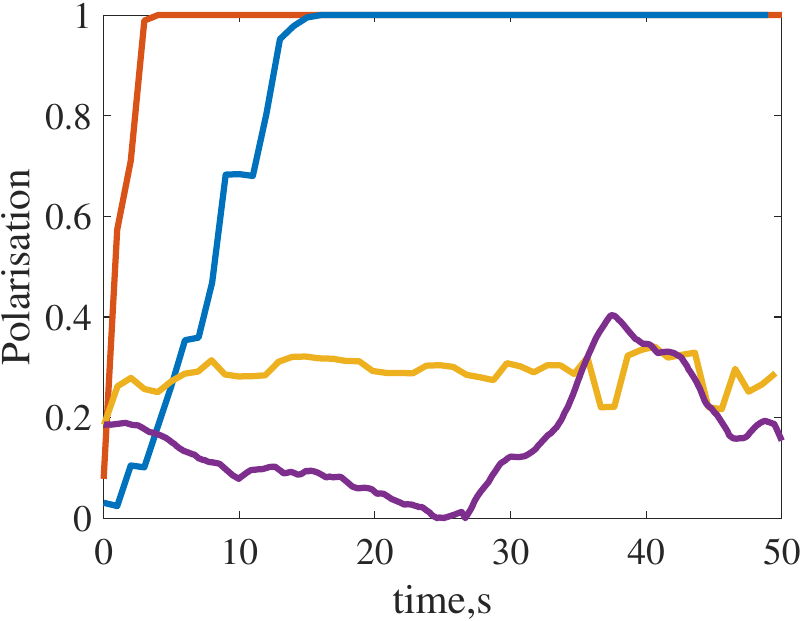}
    \label{f:conn_comp_polar}\end{subfigure} 
	\begin{subfigure}[h]{0.32\linewidth} \includegraphics[width=\textwidth]{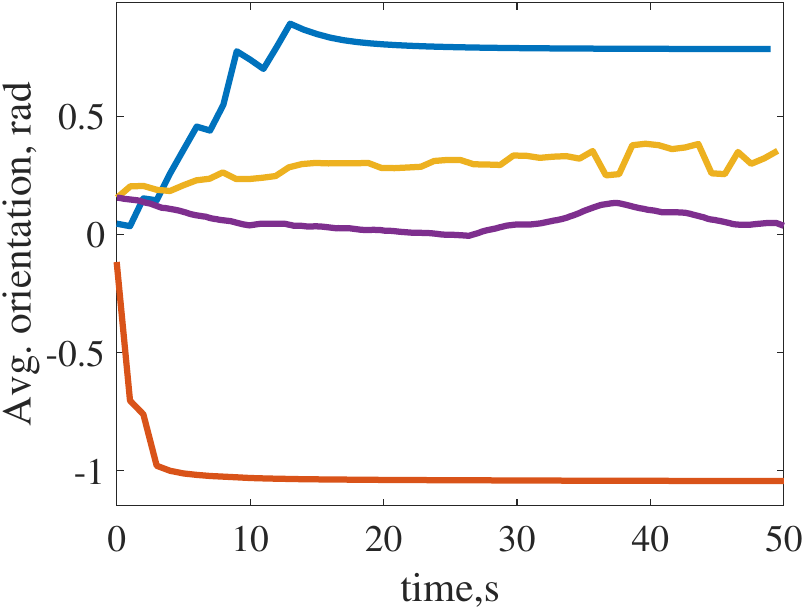}
    \label{f:conn_comp_meanOri}\end{subfigure}
    \begin{subfigure}[h]{0.32\linewidth} \includegraphics[width=\textwidth]{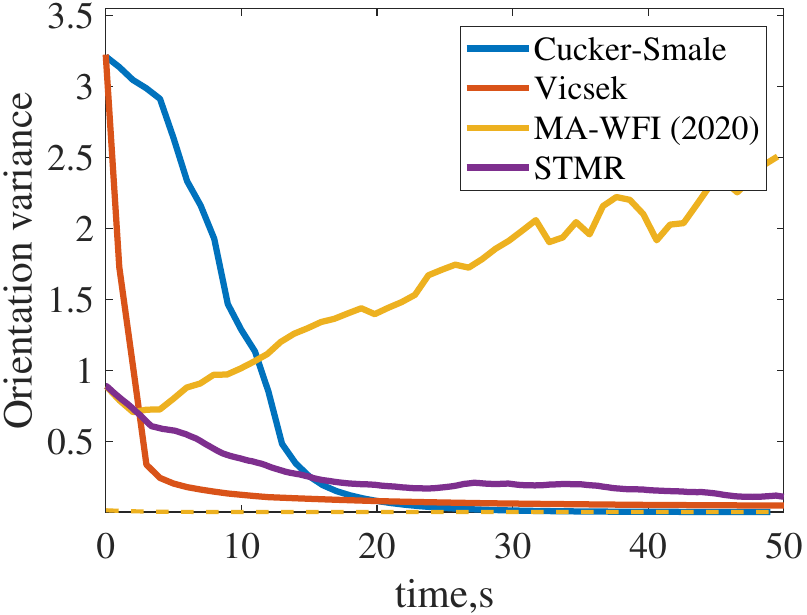}
    \label{f:conn_comp_variance}\end{subfigure} 
    \caption{Comparison of swarm polarisation (top left), average swarm heading (top right) and heading variance (bottom) over time among the STMR feedback control system proposed in this paper, the optomotor feedback control system from \citep{billah2020bioinspired}, the Vicsek model \citep{Vicsek1995} and Cucker-Smale model \citep{Cucker2007}. }
    \label{f:connectivity_compare2}
\end{figure*}

Comparison study results are shown in Figs.~\ref{f:comp_a}-\ref{f:connectivity_compare2}. Fig.~\ref{f:comp_a} shows the output of the first case where only one agent is controlled among an established swarm motion. MA-WFI agent reaches an equilibrium condition of $\theta^*=0$ from its initial orientation offset, as predicted. The Vicsek agent also reaches equilibrium, using the average orientation of its neighbors as feedback, with a response time faster than MA-WFI. The STMR agent initially continues in its initial orientation, before returning towards the other agents. It does not reach $\theta^*=0$ during the simulation but shows it can successfully participate in an established group formation. The C-S model does not participate in the motion; given how it is meant to velocity-match all agents, the model may be less applicable when only one agent is being controlled.  

Fig.~\ref{f:compare} shows the trajectory of agents for the second case where all agents are being controlled by the different swarm models. STMR agents show cohesion and group motion, as expected. MA-WFI (Fig.~\ref{f:compare}, top) was designed for a single feedback-controlled agent, and does not show convergence when all agents are feedback-controlled. Vicsek agents (Fig.~\ref{f:compare}, middle) converge to the average orientation of the group, whereas the C-S agents (Fig.~\ref{f:compare}, bottom) converge to an equilibrium velocity. Once the agents converge to their equilibrium conditions, the formation of motion does not change.

 Fig.~\ref{f:connectivity_compare} shows the connectivity plots for the four multi-agent approaches considered, with 50 agents in each simulation. Fig.~\ref{f:conn_comp_a} shows the instantaneous connectivity of the underlying undirected graph of each system, and Fig.~\ref{f:conn_comp_b} shows the connectivity of the union of underlying graphs over time. At each instant in Fig.~\ref{f:conn_comp_a}, STMR has the lowest connectivity (zero connectivity) since each agent is only connected to one other agent, whereas the other systems are either fully connected (C-S and MA-WFI) or a subset of agents are fully connected (Vicsek). In Fig.~\ref{f:conn_comp_b}, the area under the union of underlying graphs gives a measure of the total information shared (or ``work done") by the connected agents. One can quantitatively compare this effect by defining the area under the union of underlying graphs to be \textit{attentional work}. The STMR approach achieves similar attentional work as the other approaches with very low instantaneous connectivity.

Fig.~\ref{f:connectivity_compare2} shows the swarm polarisation (Fig.~\ref{f:connectivity_compare2}, top left), average swarm heading (Fig.~~\ref{f:connectivity_compare2}, top right ) and heading variance plots (Fig.~~\ref{f:connectivity_compare2}, bottom) for the simulation results from Fig.~\ref{f:connectivity_compare}. The Vicsek and C-S agents reach convergence to an arbitrary general swarm heading, resulting in swarm polarisation reaching 1 and orientation variance reaching close to 0. The MA-WFI agents show low polarisation and high orientation variance over time, indicating they do not reach convergence, which may result from all agents being controlled. The STMR agents' swarm polarisation values are continuously variable due to their `milling about' motions, and have a gradually decreasing heading variance, which indicates the agents' convergence towards a time-varying average swarm heading (bulk convection) with bounded individual orientation. 
 
\subsection{Discussion} \label{ssec:discussion}
The implementation results of the feedback control system proposed in Fig.~\ref{f:system_model} suggest that group motion can be achieved using the small-target motion reactive (STMR) framework. The agents' states are bounded over time (Fig.~\ref{sim1a}, \ref{sim1b}) and group motion is achieved. The average dwell time constraint ensures switching between agents is sufficiently slow to ensure boundedness, but the group shows general directional consensus without the constraint as well (Fig.~\ref{sim1c}). In this framework, each agent tracks only one other agent in the environment, instead of tracking all or a large subset of agents as in other traditional multi-agent systems \citep{Vicsek1995,Cucker2007}. Tracking only one neighboring agent allows the system to be easily scalable to large groups of agents without the need for increased sensory capabilities or communication bandwidth. 

The simulation results of the multi-agent system verifies that the addition of average dwell time bounds on the switching system ensures stability in terms of bounded states and prevents high frequency switching between the subsystems. Trajectories without the average-dwell time bound (Fig.~\ref{sim1c}) suggest very high frequency switching can lead to factions splitting from the group. Parameters such as $\epsilon$ and $\mu(k)$ can be used to tune the average dwell time bound to allow higher switching frequency at the cost of the trajectories converging to a bigger compact set of solutions. The trade-off is between switching frequency and robustness of the system as explained in \citep{veer2019switched}. Several mechanisms for attention-like processes and selective attentional capabilities have been discovered from studies of insect attention \citep{nityananda2016attentionAnaloguesInInsects}. The STMD neurons in insects such as dragonflies shows similar behavior by tracking only one target when multiple targets are presented \citep{wiederman2013selective}, even ``locking onto" a low-contrast target in spite of the introduction of a high-contrast distractor \citep{lancer2019target}. This ``lock-on" mechanism reduces the number of switches between targets, but does not eliminate them completely \citep{evans2022dragonfly}. While the ``lock-on" phenomenon is claimed to reduce confusion during prey selection in a swarm \citep{lancer2019target}, the results from this study may be used to show that it may also be applicable to cohesive motion coordination.

From the comparison study in Fig.~\ref{f:compare}, the STMR agent trajectories show a more unstructured behavior over time compared to the other models, due to the fact that the stability analysis only guarantees boundedness for the STMR agents instead of convergence. Factions of agents can split and rejoin (Fig.~\ref{sim1a}, bottom right), or agents show ``milling" motion \citep{hindes2021critical} that does not eventually settle down to any particular group formation. This propagation of `erratic' behavior is different than traditional swarm models found in literature that use attraction, repulsion and/or state alignment for stability. However, the lower group polarisation levels observed in STMR agent trajectories relative to C-S and Vicsek simulations (Fig.~\ref{f:connectivity_compare2}, top left) are consistent with insect swarms (e.g., honeybees, dragonflies) that have less obviously structured coordination \citep{schultz2008mechanism}.

Conventional swarming approaches often assume continuous large bandwidth of information sharing among agents (Fig.~\ref{f:connectivity_compare}), and low-error communication systems which are not easily scalable or implementable in large groups. These approaches do not account for the neural/computational limit of the biological animals that it models, nor of the hardware that will be used for implementation. Since STMR agents only need information from the largest optic flow signal producer, individual agents can be disconnected from the rest of the $N-2$ agents and still have group motion, as this study suggests (Fig.~\ref{f:connectivity_compare}). Since the information shared in conventional approaches is to provide convergence, the connectivity is much higher at the start of the simulation and decreases over time as agents converge to an alignment consensus (Fig.~\ref{f:conn_comp_a}).  Instead of having very high connectivity at the start of the simulation to prioritize convergence, STMR spreads the attentional workload over time to reach bounded trajectories instead, and shows consensus over group directionality (Fig.~\ref{f:conn_comp_b}). Low connectivity could sometimes lead to factions splitting off, but results suggest that these factions often rejoin, given the average dwell time bounds are enforced (Fig.~\ref{sim1a}, bottom left, bottom right).    

The STMD neurons in predatory insects such as dragonflies predominantly aid in pursuer-evader scenarios, STMD neurons have been found in non-predatory insects as well \citep{nordstrom2006small}. This study proposes a method of using pairwise interaction between agents for collective motion in multi-agent systems and there is precedent to believe that this might also be the case in animals. Flocking starlings interact with 6-7 neighbors \citep{ballerini2008interaction}, while honeybees in convective group milling interact with 1-2 neighbors \citep{islamFaruquePidLock}. A similar behavior as observed from STMR agents, such as faction splitting and rejoining resulting in robust cohesion, is also observed in flocking starlings to avoid predators, as the swarm constantly changes its density and structure and yet no agent is isolated from the group \citep{ballerini2008interaction}.  

\subsection{Discussion of Assumptions and Implications}
This model assumes the small-target motion detectors can reliably suppress background motion, which is generally slow-moving and further away, generating low frequency optic flow signals, similar to how STMD neurons in dragonflies respond to background motion \citep{evans2022dragonfly}. In contrast, the fast-moving small-target agents closer to the observer are assumed to provide high frequency optic flow signals that are used as input for the system. The model also assumes STMD neurons are ideal sensors with no temporal or frequency dependence, resulting in zero time delay and signal amplification over a wide range of frequencies. This assumption allows for investigating the effect of concise STMD-type neural sensing mechanisms in multi-agent systems in idealized conditions before further analysis of the sensory model including temporal dynamics.  

No exclusive collision avoidance algorithm is implemented for the STMR system, and the results suggest the STMR feedback rule does not provide repulsion in close range. In simulation, the agents were allowed to overlap and continue in their trajectories, while in experiment the agents collided and moved together after collision. Separating the effect of collision avoidance ensured the effect of STMD-based sensing in group cohesion could be analyzed in depth. 

The \citet{veer2019switched} theorem used for stability analysis of the switched system provides a bound on the state trajectories for multiple equilibrium of the sub-systems. The behaviors discussed thus require some switch rate limit provided by the average dwell time constraint. Similar behavior can be seen without the constraint (Fig.~\ref{sim1c}) but the agents are then more prone to splitting into multiple groups. Attention in insects have been well studied and several mechanisms identified for attention-like processes \citep{nityananda2016attentionAnaloguesInInsects}. Insects such as \textit{Drosophila} have also been found to have attention-like mechanisms where they can shift the focus of attention (FoA) internally, with the time-span between each shift of FoA being 1s-4s \citep{koenig2016vision}, values comparable to the average dwell time bounds used in the bio-inspired feedback model designed in this study. In addition, insect attention has multiple definitions and unifying them is outside the scope of this study. Instead, we use the ad-hoc term \textit{attentional work} to describe the area under the union of graphs (Fig.~\ref{f:conn_comp_b}), to provide a quantitative comparison of the total information shared (or ``informational work done") by the connected agents.   

The comparison and connectivity plots among different multi-agent systems in Figures~\ref{f:comp_a} - \ref{f:connectivity_compare} are one instance of simulation amid numerous other possible trajectories based on the gains and tunable parameters chosen for each system. Instantaneous connectivity values are dependent on the gains used for convergence, but the trend remains similar to that shown in Fig.~\ref{f:connectivity_compare}. Different gains might affect the time it takes for attentional work to reach equality for STMR and the other multi-agent systems, but does not necessarily affect the existence of such a finite time. 
 
\subsection{Summary and Conclusion} \label{sec:summary}
This study introduces the small-target motion reactive (STMR) approach to insect swarming by considering an engineering model of the STMD neurons found in the lobula of insects such as hoverflies and dragonflies. The two major outputs from this group of motion detection neurons are the highest magnitude of optic flow and the bearing angle corresponding to the moving agent generating the highest magnitude of optic flow. Assuming an idealized case of the STMD neuron model, a multi-agent switched feedback system framework (STMR) is developed. Stability analysis of the multi-agent system involves bi-agent stability of each subsystem, which extends to state boundedness via the Veer-Poulakakis theorem for switched systems. An average dwell time constraint ensures the agent states are bounded over time by ensuring that switching between alternating agents is sufficiently slow. The feedback control system is implemented in simulation and on ground vehicles for validation, resulting in agent trajectories that demonstrate group cohesion and general directional consensus. Comparison with other multi-agent systems such as Vicsek and Cucker-Smale models suggest that the low connectivity network with an agent only tracking one other agent may still result in group motion comparable to conventional swarm algorithms but with consistently very low sensory computational requirements. Overall, STMR represents a bio-inspired strategy that could provide a similar function to traditional engineered swarm approaches while accounting for low computational capabilities of insect visuomotor systems by spreading the processing workload over time and limiting the amount of instantaneous individual ``attention" requirements to the surrounding.    

\section*{Acknowledgment}
This work was supported in part by ONR Young Investigator Award N00014-19-1-2216. 

\section*{Research Data}
The data that support the findings of this study are available from the authors upon reasonable request.


\nolinenumbers

\bibliography{main}
\bibliographystyle{IEEEtranN} 

\end{document}